# Concerning the Superconducting Gap Symmetry in $YBa_2Cu_3O_{7-\delta}$, $YBa_2Cu_4O_8$ and $La_{2-x}Sr_xCuO_4$ Determined from Muon Spin Rotation in Mixed States of Crystals and Powders

**Dale R Harshman**[1,2,3] **and Anthony T Fiory**[4]

[1]Physikon Research Corporation, Lynden, WA 98264, USA
[2]Department of Physics, University of Notre Dame, Notre Dame, IN 46556, USA
[3]Department of Physics, Arizona State University, Tempe, AZ 85287, USA
[4]Department of Physics, New Jersey Institute of Technology, Newark, NJ 07102, USA

Email: drh@physikon.net

**Abstract**

Muon spin rotation ($\mu^+$SR) measurements of square-root second moments of local magnetic fields σ in superconducting mixed states, as published for oriented crystals and powder samples of $YBa_2Cu_3O_{7-\delta}$ (δ ≈ 0.05), $YBa_2Cu_4O_8$ and $La_{2-x}Sr_xCuO_4$ (x ~ 0.15–0.17), are subjected to comparative analysis for superconducting gap symmetry. For oriented crystals it is shown that anomalous dependences of σ on temperature T and applied field H, as-measured and extracted *a*- and *b*-axial components, are attributable to fluxon depinning and disorder that obscure the intrinsic character of the superconducting penetration depth. Random averages derived from oriented-crystal data differ markedly from corresponding non-oriented powders, owing to weaker influence of pinning in high-quality crystals. Related indicators for pinning perturbations such as non-monotonic H dependence of σ, irreproducible data and strong H dependence of apparent transition temperatures are also evident. Strong intrinsic pinning suppresses thermal anomalies in *c*–axis components of σ, which reflect nodeless gap symmetries in $YBa_2Cu_3O_{7-\delta}$ and $YBa_2Cu_4O_8$. For $YBa_2Cu_3O_{7-\delta}$, the crystal (*a-b* components, corrected for depinning) and powder data all reflect a nodeless gap (however, *a-b* symmetries remain unresolved for crystalline $YBa_2Cu_4O_8$ and $La_{1.83}Sr_{0.17}CuO_4$). Inconsistencies contained in multiple and noded gap interpretations of crystal data, and observed differences between bulk $\mu^+$SR and surface-sensitive measurements are discussed.

## 1. Introduction

The muon spin rotation ($\mu^+$SR) technique [1] is used to probe the static and dynamic characteristics of the internal magnetic field distribution, denoted as P(B), of a given material. In the case of time-differential $\mu^+$SR (the technique highlighted herein), the internal field distribution is determined by measuring the depolarization over time of an implanted $\mu^+$ spin ensemble. For type II superconductors in the mixed (fluxon) state, this technique probes the internal field distribution of the sample due to the formation of fluxons in an applied field H and thus (in the absence of electronic moments and muon diffusion) senses only the effects of the superconducting condensate. If the fluxons are static (e.g. strongly pinned), the second moment of the field distribution $\langle|\Delta B(T,H)|^2\rangle$ can yield the temperature T and field dependence of the magnetic penetration depth $\lambda(T,H)$, which directly reflect the pairing-state symmetry of the superconducting gap. For example, within the London model for a static triangular lattice of aligned fluxons, $\langle|\Delta B|^2\rangle = 0.00371\phi_o^2\lambda^{-4}$ for B << $H_{c2}$, where $\phi_o$ (= $2.068 \times 10^{-7}$ G cm$^2$) is the magnetic flux quantum [2] and $H_{c2}$ is the upper critical field. If, however, dynamic effects such as fluxon depinning and motion are present or, in the case of non-optimal high-$T_C$ compounds where inhomogeneities can occur, the intrinsic relationship between $\langle|\Delta B|^2\rangle$ and $\lambda$ (e.g. $\langle|\Delta B|^2\rangle \propto \phi_o^2\lambda^{-4}$) can be rendered invalid and the true character of the pairing state can be obscured or masked [convention in $\mu^+$SR expresses square-root variance in P(B) by the parameter $\sigma \equiv \gamma_\mu\langle|\Delta B|^2\rangle^{\frac{1}{2}}$, where $\gamma_\mu$ is the muon gyromagnetic ratio (= $2\pi \times 13.55$ MHz/kG)]. Thus to determine the temperature dependence of the penetration depth $\lambda$ (T, H→0), which directly reflects the gap function, one need only know the unperturbed second moment $\langle|\Delta B(T,H\to 0)|^2\rangle$. This is an important point, since in this work we will be considering $\mu^+$SR measurements on randomly oriented powders as well as crystals, where data for $\langle|\Delta B|^2\rangle$ are obtained from time-dependent fitting functions that appropriately model the form of P(B) [2]. For random powders a Gaussian form $\exp(-\sigma_g^2 t^2/2)$ often suffices, yielding $\sigma = \sigma_g$ and $\langle|\Delta B|^2\rangle = \gamma_\mu^{-2}\sigma_g^2$.

Early $\mu^+$SR measurements of $\langle|\Delta B(T)|^2\rangle$ acquired on sintered powder [3,4] and twinned single-crystal [5] samples of YBa$_2$Cu$_3$O$_{7-\delta}$ found temperature dependences obeying the form $\lambda^{-2}(T) \propto 1 - (T/T_C)^\alpha$, where the exponent $\alpha \approx 4$ showed agreement with the Gorter and Casimir two-fluid model (for superconducting transition temperature $T_C$) [6-12]. The interpretation proposed for these results is that the gap symmetry of the bulk superconducting state is nodeless and consistent with strong coupled *s*-wave pairing. Results for the *a-b* basal-plane component $\lambda_{ab}(0)$ among the three experiments are also in close agreement [13]. One learns from these early studies that valuable information can be obtained from data on high-quality powders (grain sizes typically on the order of a few $\mu$m >> $\lambda$), which must be considered in deducing a valid understanding of $\mu^+$SR measurements in the high-$T_C$ compounds. Later studies of high-purity YBa$_2$Cu$_3$O$_{6.95}$ crystals produced conflicting results. The $\mu^+$SR data indicated a linear-like decrease in $\lambda^{-2}(T)$ at low T [14], and analysis of microwave surface impedance measurements, which provided greatly improved resolution, produced a temperature dependence $\Delta\lambda(T) \propto T$ (absolute $\lambda$ undetermined) [15,16]; both results were proposed by the authors as evidence of a noded (*d*-wave) gap symmetry (contradictions noted in section 2). Interestingly, the development of variable-energy slow-$\mu^+$ beams [17] has recently enabled $\mu^+$SR to profile penetration of weak magnetic fields at the surface of YBa$_2$Cu$_3$O$_{6.92}$ (state-of-the-art high quality crystals) [18]. It was found that a dead surface layer (about 10% of $\lambda$) was needed to model $\lambda$ with Meissner-state theory (similar anomalies were found in previous studies [19,20]). These low-energy $\mu^+$SR profiling results indicate that interpretation of surface-specific measurements of $\lambda$ is less direct than presumed (as discussed further in section 5.4), making the understanding of bulk $\mu^+$SR measurements of $\lambda$ all that more imperative. Numerous surface or interface specific experiments on YBa$_2$Cu$_3$O$_{7-\delta}$ crystals show clear evidence for gap nodes, and specifically *d*-wave pairing (e.g. tunnelling and frustrated junctions [21] and angle-resolved photoemission spectroscopy [22]) although the *d*-wave picture fails to account for residual electrodynamic conductivity (infrared and microwave) [23] and some scanning tunnelling microscopy (STM) results [24]. The objective of this work is to analyze $\mu^+$SR measurements of $\sigma(T,H)$ in the mixed state, for the purpose of determining the intrinsic $\lambda(T,H)$ and thereby the true nature of the gap in the bulk of the superconducting state (as opposed to at surfaces or interfaces). We discuss the utility of σ(T,H) data from crystal and powder specimens in providing this essential piece of information.

It was previously shown that temperature-activated fluxon depinning can introduce inflections and linear-like features in the temperature and field-dependence of the *a-b* planar component $\sigma_{ab}$(T,H) for single-crystal $YBa_2Cu_3O_{7-\delta}$ [25,26]; of interest here is that the intrinsic form of $\lambda_{ab}$(T,H) can be extracted by modelling the pinning perturbation. Unambiguous indicators of fluxon depinning effects are:

1. σ(T,H) differs qualitatively among oriented crystals and from non-oriented sintered powders.

2. σ measurements repeated on the same crystal and apparatus give differing results.

3. σ is non-monotonic with applied field H, particularly as T→0.

4. $T_C$ appears to decrease with applied field for H << $H_{C2}$(T).

Whenever any of these indicators are present, or if the material is non-optimal, then it is necessary to develop a theoretical approach that takes such phenomena into account. For weak pinning in non-optimal high-$T_C$ compounds, fluxon effects (i.e. depinning, motion) would likely dominate, while in strong pinning cases (such as in non-oriented sintered powders) the effects of quasi-particle scattering might become more important. Anomalous temperature (and field) dependences in σ(T,H) can also be introduced by sample inhomogeneity (typically arising from off-stoichiometric disorder in the high-$T_C$ cuprates), which can also mimic some of the characteristics of higher-order pairing. In non-optimal compounds with depressed $T_C$, such as underdoped (non-oxygen-ordered phase) $YBa_2Cu_3O_{7-\delta}$ or $La_{2-x}Sr_xCuO_4$, or perhaps $YBa_2Cu_4O_8$ at ambient pressure, enhanced quasi-particle scattering may be modelled by multiplying $\lambda^2$(T) by the factor [1+ $\xi/\ell$(T)], which is a function of an effective coherence distance $\xi$ and a temperature-dependent mean free path $\ell$(T), e.g. $\ell^{-1}$(T) containing a term linear in T as extrapolated from the normal state. This restricted mean free path reduces negative curvature in σ(T) and produces a more linear-like temperature dependence.

An examination of $\mu^+$SR results obtained from powder and crystalline forms of various superconducting compounds are presented in section 2; the importance of including comparisons between the two material forms in a comprehensive analysis is demonstrated. Single-crystal data, in particular, show self-evident sample-dependent differences in temperature dependence. Section 3 describes the necessary components of a theoretical description of temperature-dependent fluxon pinning for interpreting data. In section 4 we compare the temperature- and field-dependences of σ(T,H) from $\mu^+$SR experiments on oriented crystals of $YBa_2Cu_3O_{7-\delta}$, $YBa_2Cu_4O_8$ and $La_{1.83}Sr_{0.17}CuO_4$ and corresponding random powder samples of this family of compounds, noting that marked differences between the two sample forms are observed. These experiments are particularly instructive in elucidating the obvious extrinsic effects. We show that a rational analysis based on temperature-activated fluxon depinning, coupled with sample inhomogeneity in some cases, provides a self-consistent explanation for the anomalous behaviour observed in crystals, owing to their weaker pinning, and the suppression of such anomalies in sintered powders, with comparatively stronger pinning. Model corrections for depinning are used to obtain relative probabilities on whether intrinsic penetration depths correspond to noded (*d*-wave) or nodeless [two-fluid and Bardeen-Cooper-Schrieffer (BCS)] gap symmetries from analysis of 15 sets of data on crystals and powder specimens. These results are discussed in detail in section 5, and our conclusions are presented in section 6.

## 2. $\mu^+$SR in crystals and powders

Good examples of a static fluxon (or vortex) lattice are found in the early $\mu^+$SR results for the two sintered powder [3,4][5] and a mosaic of (slightly non-optimal) twinned single-crystal [5] samples of YBa$_2$Cu$_3$O$_{7-\delta}$ (with the crystals having $\delta \sim 0.1$, transition temperature T$_C$ = 82 K, width $\Delta$T$_C$ < 0.7 K from magnetization), given that two-fluid forms are consistently found in the temperature dependence of $\langle|\Delta B(T)|^2\rangle$. Deviations from this response in the high-T$_C$ compounds can arise from two possible sources: (1) sample inhomogeneity due to non-optimal stoichiometry [27] or (2) temperature-activated fluxon motion and depinning [28]. For the early YBa$_2$Cu$_3$O$_{7-\delta}$ samples studied [3-5], the fluxons were strongly pinned for all T ≤ T$_C$, as exhibited by the pinned form of P(B) in sintered powders [29], thereby negating the possibility of perturbations in the temperature dependence of $\langle|\Delta B|^2\rangle$ arising from fluxon motional effects, which might obscure the true nature of the gap function. A small deviation from a nodeless strong-coupling character (e.g. $\alpha$ < 4) was, however, observed for an overdoped, non-oriented sintered powder sample of Bi$_2$Sr$_2$CaCu$_2$O$_{8+\delta}$ (T$_C$ = 72 K) [30]. In this case, measurements of the Gaussian relaxation rate of the muon spin polarization $\sigma_g$(T) conducted in an applied field of H = 1 T exhibited a slightly stronger T-dependence than that expected for a strong-coupled, s-wave gap [30], which was attributed to fluxon dynamics (herein we use SI units with $\mu_0$ implicit). Given that Bi$_2$Sr$_2$CaCu$_2$O$_{8+\delta}$ (with effective mass anisotropy $\Gamma \equiv m^*_c/m^*_{ab}$ > 1300 – 3000 [31,32]) is far more anisotropic than YBa$_2$Cu$_3$O$_{7-\delta}$ ($\Gamma \geq 25$ [5]), it is not surprising that significant fluxon motion is observed even in sintered powder samples of this material [33,34].

---

[5] Both were high quality powder specimens exhibiting near 100% Meissner effect and sharp transitions.

Clear and unambiguous evidence from $\mu^+$SR of fluxon depinning in the high-T$_C$ compounds was first reported for single-crystal (unannealed) samples of Bi$_2$Sr$_2$CaCu$_2$O$_{8+\delta}$, as indicated by the temperature and field dependence of $\langle|\Delta B(T,H)|^2\rangle$ for magnetic fields applied perpendicular to the *ab*-plane [28]. For intermediate fields, H = 0.3 and 0.4 T, $\langle|\Delta B(T,H)|^2\rangle$ exhibits an inflection at T ≈ 15–20 K, and an extrapolated value for T→0 that decreases with applied field H. At higher applied fields (e.g. 1.5 T) the inflection vanishes, and $\langle|\Delta B(T,H)|^2\rangle$ recovers a more conventional (i.e. s-wave-like) T-dependence (but with a substantially suppressed magnitude as T→0 compared to that measured at intermediate fields). Interpretation of these data, which has withstood the test of time, is that temperature-activated fluxon depinning occurs at the intermediate fields (0.3 and 0.4 T), as identified by the inflection point. At 1.5 T, fluxon motion is minimized, accompanied by a cutoff in the c-axis correlation length of the flux lines [35] (e.g. misalignment of fluxons between superconducting layers), the latter being responsible for the precipitous decrease in $\langle|\Delta B(T\to 0,H)|^2\rangle$ with increasing field [35]. The high quality of the crystals, which have fewer defects as compared to powder samples [30], combined with the large $\Gamma$ of Bi$_2$Sr$_2$CaCu$_2$O$_{8+\delta}$, allows the temperature-activated fluxon depinning to dominate the temperature and field dependences of $\langle|\Delta B|^2\rangle$ [35].

Similar anomalous temperature and field dependences of $\langle|\Delta B|^2\rangle$ were eventually observed in full line shape analyses of P(B) data for YBa$_2$Cu$_3$O$_{7-\delta}$ crystals, as the quality of specimens improved over time with advancements in crystal growth [14,25,26,36,37]. Initially, linear-in-T forms were seen at low temperatures that were deemed as evidence of *d*-wave pairing under the rationale that the $\mu^+$SR data resemble a similar linearity in the $\Delta\lambda$ derived from microwave impedance [14,36]. The later studies revealed the more dramatic anomaly of inflections in the temperature dependence near 10–15 K [25,26,37]. An illustration of differences in results for two better quality crystals is presented in figure 1, where YBCO1 denotes the initial $\lambda_{ab}$ data (at fields H = 0.2 and 0.5 T) in [36] and YBCO2 denotes latest $\sigma_{ab}$ data (H = 0.1 T) in [37] (plot scales correspond to $\sigma_{ab} = (0.00371)^{1/2}\gamma_\mu\phi_o\lambda_{ab}^{-2}$, for *a-b* basal-plane components [13]). The upper continuous curve is the fit from *d*-wave theory, which includes non-local, non-linear and vortex-core effects, to YBCO1 data for H = 0.5 T as derived in [38], where it was observed that the theory does not agree with the data (*d*-wave theory is distinctly curved, whereas temperature dependence in the data appears linear). As noted in [38], data at low H are preferred for distinguishing the peculiar linear-T form of *d*-wave theory, because *d*- and *s*-wave forms both have temperature dependences exhibiting negative curvature at finite H. Under that criterion the lower-H data for YBCO2 [37] is in principle more appropriate for studying pairing symmetry; however, the temperature dependence turns out to be completely different from YBCO1, particularly in the much stronger temperature dependence at low temperatures. The YBCO1 crystal evidently has strong pinning, as was proved in [14] by the field-shifting technique that observes a frozen form for P(B) (qualitatively unchanged lineshape and average internal induction $\langle B \rangle$ nearly constant upon changing applied H). The temperature-dependent $\lambda_{ab}^{-2}$ for YBCO1 in figure 1 reflects the variation in P(B) with temperature, owing to temperature dependence in the intrinsic penetration depth and the extrinsic flux pinning [39]. The dashed curve in figure 1 for YBCO2 was obtained with our depinning model (data analysis and alternative two-gap interpretation of [37] are discussed in further detail in section 4).

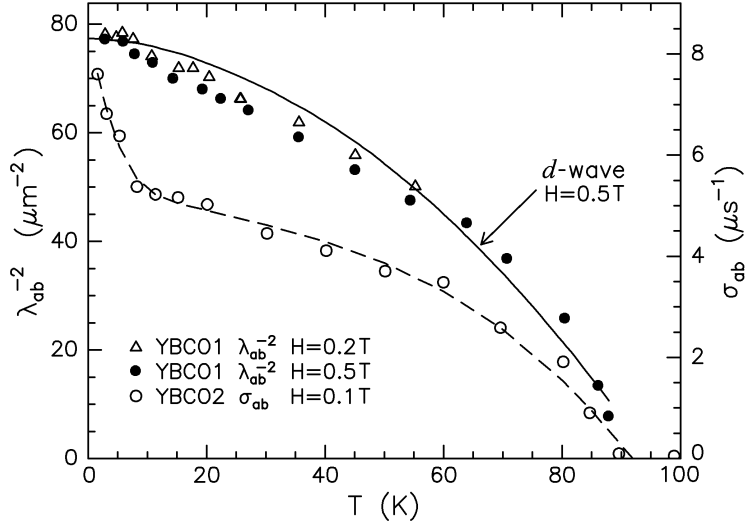

**Figure 1**. $\mu^+$SR data for two high-quality $YBa_2Cu_3O_{7-\delta}$ crystals: $\lambda_{ab}^{-2}$ (left scale) for YBCO1 in [36]; $\sigma_{ab}$ (right scale) for YBCO2 in [37] (identical to (A) of figure 3) taken at applied fields indicated. Solid curve is $d$-wave fit from [38]. Dashed curve is depinning model.

A related observation associated with pinning is strong non-monotonic dependence of $\langle |\Delta B(T,H)|^2 \rangle$ on field (particularly noticeable as $T \to 0$). As was shown earlier [25,26,39], these anomalous temperature and field dependences in $YBa_2Cu_3O_{7-\delta}$ crystals are consistent with the influence of fluxon depinning in the presence of an intrinsic $\lambda_{ab}(T,H)$ that is best described by the two-fluid formula [6-12]. These crystal results confirmed the nodeless character originally observed in powder and twinned crystal samples [3-5], where the fluxons are strongly pinned, and showed that anomalies in $\langle |\Delta B(T,H)|^2 \rangle$ are symptomatic of temperature-activated depinning of fluxons and distortions of the flux lattice.

Acting on the premise of continual improvements in crystal preparation and quality over time, we consider for this study data from three recent time-differential $\mu^+$SR experiments on oriented crystals of $YBa_2Cu_3O_{7-\delta}$ (denoted YBCO2 in figure 1) [37], $YBa_2Cu_4O_8$ (at ambient pressure) [40] and (slightly overdoped) $La_{1.83}Sr_{0.17}CuO_4$ [41]. All three studies provide data for $\sigma_{ab}(T,H)$ (extracted using multiple-Gaussian fits to accurately model P(B) [42]) exhibiting an inflection in the temperature dependence near 10-15 K and a strong dependence on field (field-dependent data available for $YBa_2Cu_3O_{7-\delta}$ [37] and $La_{1.83}Sr_{0.17}CuO_4$ [41]); i.e. the presence of the aforementioned signatures of temperature-dependent fluxon depinning. In the case of $La_{1.83}Sr_{0.17}CuO_4$, quasi-particle scattering due to non-optimal stoichiometry also plays a role. The interpretation common to [37,40,41] assumes optimal compounds and a static fluxon structure in equilibrium for all $T \leq T_C$, and attributes the inflection to intrinsic superconductivity containing a $d$-wave gap and small secondary $s$-wave gap. In this work we show how temperature dependent pinning and depinning can create illusions of $d$-wave pairing and multiple gaps.

## 3. Theoretical modelling of fluxon depinning

In the following we consider a physical model that treats the distortions of the flux-line lattice caused by random local pinning forces [33,25]. One recognizes that to minimize errors from fluxon movement, $\mu^+$SR studies of the superconducting mixed state (in type II superconductors) generally employ field-cooling methods with step-wise variations in temperature at constant H. However, the equilibrium flux density theoretically does not remain constant upon cooling, owing to the temperature dependence of the diamagnetism in the mixed state. In the presence of pinning the diamagnetic screening currents can induce flux creep and spatial variations in the magnetic induction B that increase $\langle|\Delta B(T,H)|^2\rangle$ [28,43,44]. The various dynamical fluxon phases that form upon field-cooling (Bragg glass, near-vortex lattice) [26] can lead to non-monotonicity in $\langle|\Delta B(T\to 0,H)|^2\rangle$. Macroscopic manifestations of flux pinning can therefore appear in the temperature- and field-dependences of the first moment $\langle\Delta B(T,H)\rangle$ and even the third moment $\langle|\Delta B(T,H)|^3\rangle$, as has been observed [28]. Motional narrowing (which decreases $\langle|\Delta B(T,H)|^2\rangle$) is presumed to be negligible (except for $T \sim T_C$) for the lower-$\Gamma$ (La,Sr) and (Y,Ba) cuprates of interest here. To render this problem tractable we consider a quasi-static model, where fluxon motion occurs primarily during the periods of temperature changes. In comparing the model to data for $\sigma$, we also presume a diligent experimental protocol, wherein sufficient time has been allotted at a given temperature for the fluxon distribution to reach metastable equilibration prior to data collection.[6] Fluxon displacements are then treated as static distortions induced by random pinning in the fluxon array that would otherwise be a triangular lattice of straight flux lines. Below we analyze the temperature and field dependences of the square-root second moment, $\sigma(T,H) \equiv \gamma_\mu\langle|\Delta B|^2\rangle^{1/2}$ (presented herein in units of $\mu s^{-1}$). We shall show that this quasi-static pinning model, originally applied to single-crystal $YBa_2Cu_3O_{7-\delta}$ [25,26], well represents the observed behaviours of second-moment data for crystals and powders of the three compounds treated in this work.

For the convenience in defining terms, we briefly review the method for analyzing the $\sigma$ data. Following Brandt [33], the transverse fluxon displacements are expressed in terms of displacements $u_l$ of the smooth fluxon lines from their lattice sites and displacements $u_p$ of fluxon points along their smooth lines. The $u_l$ fluctuations tend to increase the root second moment of the local field distribution, while the $u_p$ fluctuations tend to decrease it. The root second moment $\sigma$ of the local magnetic field distribution in a fluxon lattice with random displacements is determined as

$$\sigma = f_p \sigma_0, \qquad (1)$$

where $\sigma_0$ is the root second moment of the unperturbed fluxon lattice. The factor $f_p$ is determined from equation (17) of [33],

$$f_p = [\exp(-26.4\, u^2/a^2) + 24.8\, (u_l^2/a^2)\ln(\tilde{\kappa})]^{1/2}, \qquad (2)$$

where $a = (2\phi_o/3^{1/2}\langle B\rangle)^{1/2}$ is the mean intervortex spacing,[7] $\langle B\rangle$ is the average magnetic induction and

$$u^2 = u_l^2 + u_p^2. \qquad (3)$$

Equation (2) contains the parameter

---

[6] Temperature changes must be sufficiently slow to avoid irreproducible results for $\sigma$ at low temperature.
[7] Assume equilateral triangular lattice approximation, oblique in $YBa_2Cu_3O_{7-\delta}$ (see STM in [24]).

$$\tilde{\kappa} = [(u_l^2 + 2\lambda^2)/(u^2 + 4\xi^2)]^{1/2}, \tag{4}$$

where $\xi^2 = \phi_0/2\pi H_{c2}$. Here we assume collective pinning by random, local pinning forces that induce deformations in the fluxon lattice. In an earlier paper, Brandt (see equation (9) of [43]) showed that the increase in the second moment of the local field distribution from compressive fluctuations scale with the pinning energy. Following [25], the pinning energy is therefore assumed to be proportional to $u_l^2$ and also temperature dependent, so one has

$$u_l^2 = u_{l0}^2 \, f_l(E_P/k_B T), \tag{5}$$

where the function $f_l(x) = \exp[-1.45x - 0.386x^2]$ approximates thermally activated depinning in the presence of a Gaussian distribution in $u_l^2$ [25], $E_p$ is a characteristic depinning energy, and $u_{l0}$ is the zero-temperature limit of $u_l$. Equation (5) expresses the thermal transition from a uniform triangular lattice ($u_l \sim 0$) in the limit of high temperature to a fluxon lattice distorted by pinning at low temperatures ($u_l \sim u_{l0}$). As in [25], equation (5) is applied by treating $E_p$ as a materials constant, i.e. independent of applied field, and $u_{l0}$ as field dependent.

Lateral fluctuations along the flux lines are modelled by the parameter $u_p$. Brandt treated this case as random lateral fluctuations of point fluxons along an otherwise smooth line [33]. In general transverse waves along the flux lines increase the line tension and entail a cost in energy. The energy is supplied by pinning forces and thermal energy, which scales with $k_B T$, such that this contribution can be modelled by a form with linear temperature dependence,

$$u_p^2 = u_{p0}^2 + u_{p1}^2 \, T/T_C. \tag{6}$$

The depinning model thus depends upon the empirical parameters, $u_{l0}$, $E_P$, $u_{p0}$, and $u_{p1}$. We note general validity in applying this analysis to second moments $\langle|\Delta B^2|\rangle$, irrespectively of the form taken by the temperature dependence of the first moment $\langle B \rangle$ (which for sufficiently strong pinning can be nearly constant, as found for fixed H in [26] and for field-shifted H in [14]), since $u_l$ and $u_p$ represent the short-range disorder in pinned vortex lattices.

According to the London model, the root second moment of the fluxon lattice, in units of the muon spin precession rate, is given by the expression [43],

$$\sigma_0 = \gamma_\mu [0.0609 \, \phi_o / \lambda^2]. \tag{7}$$

Given the earlier results for $YBa_2Cu_3O_{7-\delta}$ [3-5,25,26] as well as the experimental trends observed in sections 4.1 and 4.2 below, we examine whether the temperature dependence of the magnetic penetration depth associated with the true gap function is well approximated by the two-fluid formula [6-12],

$$\lambda^{-2}(T,H) = \lambda_L^{-2} f(H/H_{c2}(T)) [1 - (T/T_C)^4], \tag{8}$$

where $\lambda_L$ is the London value. The function $f(H/H_{c2})$, which models the field dependence (and includes the core correction), is obtained from calculations of fluxon lattice structure within quasiclassical Eilenberger theory [44] and can be expressed in an analytical form in the region of interest $H \leq 0.9\ H_{c2}$ as[8]

$$f(h) = (1 - 1.062\ h^{1/2} + 0.0776\ h)\ . \qquad (9)$$

We also examine alternative forms for $\lambda^{-2}(T,H)$ that are obtained from theory for strong-coupling BCS, taking $N(0)V = 4$ (product of density of states of one spin and pairing interaction potential) as exemplary, and for *d*-wave pairing symmetry [38], following our previous analysis of second moment data for a $YBa_2Cu_3O_{7-\delta}$ crystal [25]. Thus we consider three model functions for the intrinsic $\lambda(T,H)$, representing nodeless 2-fluid and BCS, and *d*-wave forms for the gap function, and apply the depinning model to calculate a theoretical function $\sigma(T,H)$ according to equation (1).

The model was then fitted to data for $\sigma(T,H)$ for all the specimens considered in this work using the Simplex method to adjust values for $n_p$ parameters that minimizes the $\chi^2$ statistic (normalized mean square error per degree of freedom $\upsilon = n_d - n_p$, and where $n_d$ is the number of data points). Confidence limits on the parameters were determined by the method of fitting a large number (128 in this work) of synthetic data sets, where the data in each are randomized with a standard deviation derived from measurement uncertainty [45], yielding a distribution in the values for each parameter. Standard ($\pm$) error bars for the fitted parameters are then obtained (to about 10% accuracy and exclusive of parameter correlations) from one-half of the interval containing the central 68.5% of the distribution for each parameter.

Approximating the lineshape of the mixed state by fitting multiple Gaussians to $\mu^+SR$ time series data, as was done for the crystal data [37,40,41], allows one to extract a second-moment parameter $\sigma^2$ associated with the formation and characteristics of the fluxon lattice. We recognize the potential for systematic improvement in determination of the second moment of P(B) by such techniques as Fourier transformation or, with improved signal-to-noise, by maximum entropy methodology [46].

## 4. Experimental trends and interpretations

It is important to point out that all of the single-crystal data being considered herein satisfy one or more of the four indicators of fluxon motion/depinning as given in section 1. Thus, any interpretation or analysis not explicitly including such effects is necessarily rendered at least incomplete, if not entirely incorrect. Below, we provide insights into this problem from two different perspectives: (A) By comparing the crystalline random average data to that acquired on powder samples, and applying our depinning model, we prove that the low-temperature inflection present in the three field-oriented components of $\sigma(T)$ for $YBa_2Cu_3O_{7-\delta}$ and $YBa_2Cu_4O_8$ (and by inference the *ab*-component of $La_{1.83}Sr_{0.17}CuO_4$) is not reflective of an additional gap function; and (B) We note the suppression of fluxon motion along the hard axis (i.e. the *c*-axis for the materials considered herein) in quasi-2D superconductors, such that $\sigma_c(T)$ better reflects the true gap function – virtually free of the obscuring effects of fluxon depinning. In the latter two cases we also discuss the pathologies introduced in non-optimal compounds.

*4.1. Comparison with random powder samples and application of depinning model*

For oriented crystal specimens the planar components $\sigma_{ab}(T,H)$, $\sigma_{ac}(T,H)$ and $\sigma_{bc}(T,H)$ are determined with the magnetic field applied along the primary crystallographic directions *c*, *b*, and *a*, respectively. In the data refinement method of [37] and [40] individual *a*-, *b*- and *c*-axis components of $\sigma(T,H)$ were then mathematically extracted according to the equations,

---

[8] Numerical fit, see figure 13 in [44].

$$\sigma_i = \sigma_{ij}\sigma_{ik} / \sigma_{jk}, \qquad (10)$$

where i, j, and k represent *a*, *b*, or *c*. For both YBa$_2$Cu$_3$O$_{7-\delta}$ [37] and YBa$_2$Cu$_4$O$_8$ (nonoptimal at zero applied pressure) [40] the results for $\sigma_a(T)$ and $\sigma_b(T)$ show an inflection as a function of T in the range 10–15 K, while $\sigma_c(T)$ exhibits no such feature (discussed in section 4.2.). To prove that the low-temperature inflection is unrelated to the superconducting hole condensate gap function, we construct the root-mean-square average of the three planar $\sigma_{ij}$ components measured for an oriented crystal sample [47],

$$\sigma = [(\sigma_{ab}^2 + \sigma_{bc}^2 + \sigma_{ca}^2)/3]^{1/2}. \qquad (11)$$

Now consider comparing this with an independent measurement of σ for a non-oriented powder sample of the same compound, in which case σ is isotropic with respect to applied field orientation. Any intrinsic phenomena captured by the root second moment should be qualitatively the same for both the non-oriented powder and the component average calculated for the corresponding crystalline sample.

*4.1.1. YBa$_2$Cu$_3$O$_{7-\delta}$* The $\sigma_i(T)$ components for single-crystal YBa$_2$Cu$_3$O$_{7-\delta}$ (T$_C$ = 91.2 K, ΔT$_C$ = 2 K and Γ ≈ 33 as determined in section 5.2) were extracted from planar-component data at two applied fields (H = 0.012 T along the *a* and *b* axes and 0.1 T along the *c*–axis) [37]. Since the raw data for $\sigma_{ac}(T)$ and $\sigma_{bc}(T)$ at H = 0.012 T are not provided, we calculated them as $(\sigma_a\sigma_c)^{1/2}$ and $(\sigma_b\sigma_c)^{1/2}$, respectively. Figure 2 shows the average according to equation (11) of the field-orientation crystalline data for the YBa$_2$Cu$_3$O$_{7-\delta}$ sample (triangles), compared to data from two powder samples: (A) from [4] (T$_C$ = 89.5 K, H = 0.35 T, open diamonds) and (B) from [48] (T$_C$ = 91.45(5) K, H = 0.2 T, open circles[9]). As is evident, the component average of the crystalline data retains the anomalous temperature dependence in σ(T), which is missing (within statistical error) in the powder data. The absence of an anomaly in the powder data [3,4,48] at low temperatures is also confirmed by other experiments at H = 1.8 T [5] and 0.35 T [49] (bear in mind that the inflection persists to at least 3 T in single crystals of this compound [25]). Moreover, figure 2 of [49] confirms the weak variation of σ(T) with field (H greater than ~0.1 T) for powder samples of YBa$_2$Cu$_3$O$_{7-\delta}$, YBa$_2$Cu$_4$O$_8$ and Bi$_2$Sr$_2$CaCu$_2$O$_{8+\delta}$, and no anomalous increases in σ(T) for H < 0.1 T. It is clear that none of the powder data for YBa$_2$Cu$_3$O$_{7-\delta}$ exhibit the same qualitative behaviour as the single-crystal component average, thus satisfying the first indicator of fluxon depinning effects listed in section 1.

Consequently, the data (single-crystal and powder) in figure 2 have been fitted using our depinning model described above assuming three possible underlying pairing states: Two-fluid, strong-coupling BCS [N(0)V = 4] and *d*-wave. For the single crystal the model was fitted to the data for the three planar components (with a common value for E$_p$). The resulting fitted parameters and $\chi^2$ statistics are given in table 1(a) for the single crystal and table 1(b) for the two powders. Best fits (smallest $\chi^2$) are obtained with the two-fluid model; probabilities that BCS or *d*-wave models (each with larger $\chi^2$) would yield comparably good fits, denoted as P(F-test), are determined from F-distribution analysis of $\chi^2$ ratios, which are normalization independent [45]. The low probabilities obtained with the two alternative models convey a high degree of confidence that the two-fluid model correctly describes the data. The fitted depinning model with the two-fluid gap function [equation (8)] is represented by the solid and dashed curves in figure 2 [components of the single-crystal fits combined according to equation (11)].

---

[9] Optimum-doped powder in oxygen isotope-effect study.

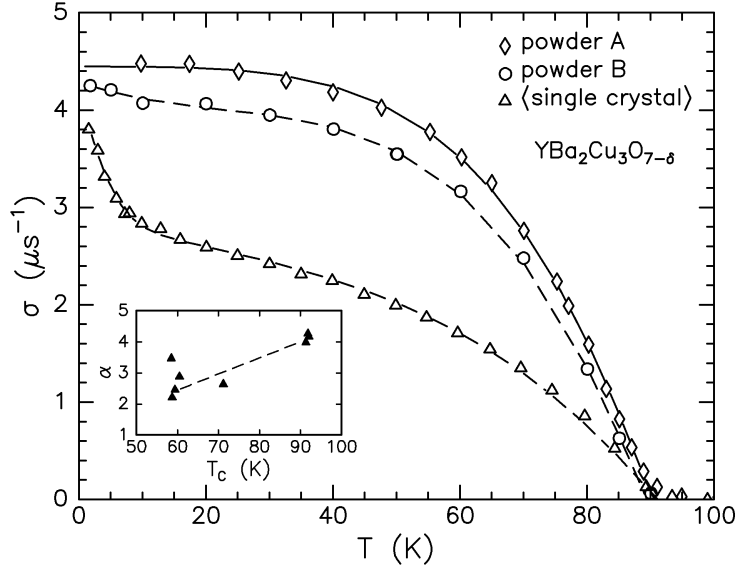

**Figure 2.** Root second moment data for $YBa_2Cu_3O_{7-\delta}$ specimens. The triangles represent the average of the single crystal field-oriented data from [37] according to equation (11). The diamonds and circles correspond to the non-oriented powder data from "A" [4] (H = 0.35 T) and "B" [48] (H = 0.2 T), respectively. The solid and dashed curves through the data represent fits of the depinning model assuming a two-fluid form for the gap function (fitted parameters in tables 1(a) and 1(b)). Inset: Plot of fitted exponent α in equation (12) versus $T_C$ for oxygen-deficient $YBa_2Cu_3O_{7-\delta}$ (data taken from [49]). The dashed line denotes the trend.

Another revealing feature of the $YBa_2Cu_3O_{7-\delta}$ data presented in [37] is the significant difference between the two sets of $\sigma_{ab}(T)$ data provided (i.e. the second indicator of fluxon depinning effects presented in section 1). These data, determined from figures 1 and 3 of [37] and denoted as measurements A and B, respectively, are compared in figure 3 (A is same as YBCO2 in figure 1). Since both data sets were acquired on the same sample, at the same magnetic field and over the same temperature range, one would expect them to coincide. While the difference cannot be intrinsic, it can be readily attributed to fluxon pinning effects which can vary with the data acquisition methodology (the glassy relaxation of the fluxons is affected by the temperature scan times or direction and is reflected in the measured σ(T,H); see the (T,H) phase diagram in [26]).

Table 1(c) presents the results of fitting the data sets of measurements A and B with the three models for the gap function introduced above, treating $E_p$ and $\lambda_{ab}(0)$ as global parameters. One again finds that the two-fluid gap model provides the best fit and it is shown by the curves in figure 3.

Upon careful examination the results of the multiple gap analysis of data for single crystal $YBa_2Cu_3O_{7-\delta}$ presented in [37] also reveal the influence of pinning in the data. In principle each intrinsic component of $\sigma_{ab}(T\to 0, H)$ ought to be a monotonically decreasing function of the applied field (in the region of measurement) [44]. In fact both components, denoted as $\sigma^d(0)$ and $\sigma^s(0)$, exhibit a peak in the field dependence (at H ~ 0.1 – 0.2 T) well outside the error bars (see figure 3, inset). (Note that the labels for these two components were mistakenly interchanged in table I of [37].) Thus $\sigma_{ab}(T\to 0, H)$ is non-monotonic with H, which satisfies the third listed indicator of fluxon-pinning effects above; the inflection in σ(T) is, therefore, associated with temperature-activated fluxon depinning [25,26]. A similar but less pronounced maximum in σ(H) at H = 0.1 T is observed in powder samples, for which pinning distortion of the fluxon lattice is essentially frozen over most of the temperature range [4].

**Table 1.** (a) Depinning model fit results for single-crystal YBa$_2$Cu$_3$O$_{7-\delta}$ (T$_C$ fixed at 91.2 K). F-test probabilities are relative to best-fitting model (two-fluid). Data from [37]. (b). Depinning model fit results for randomly-oriented powder samples of YBa$_2$Cu$_3$O$_{7-\delta}$. (c) Depinning model fit results for single-crystal YBa$_2$Cu$_3$O$_{7-\delta}$ (data sets A and B in figure 3, H = 0.1 T, T$_C$ = 91.2 K, $\upsilon$ = 31). Data from [37].

| (a) | Component | Fitting parameter | Two-fluid | N(0)V = 4 | d-wave |
|---|---|---|---|---|---|
| | ab-plane | $\lambda_{ab}(0)$ (nm) | 119(6) | 158.9(3) | 119(5) |
| | H = 0.1 T | u$_{p0}$/a | 0.18(2) | 0.000(0) | 0.17(3) |
| | | u$_{p1}$/a | 0.191(1) | 0.02(1) | 0.013(2) |
| | | u$_I$/a | 0.128(9) | 0.230(1) | 0.13(1) |
| | bc-plane | $\lambda_{bc}(0)$ (nm) | 319(2) | 330.9(9) | 119(1) |
| | H = 0.012 T | u$_{p0}$/a | 0.021(1) | 0.006(6) | 0.378(2) |
| | | u$_{p1}$/a | 0.148(6) | 0.005(2) | 0.030(4) |
| | | u$_I$/a | 0.151(9) | 0.170(5) | 0.59(20) |
| | ac-plane | $\lambda_{ac}(0)$ (nm) | 323(4) | 365(2) | 330(21) |
| | H = 0.012 T | u$_{p0}$/a | 0.12(2) | 0.029(7) | 0.06(3) |
| | | u$_{p1}$/a | 0.151(6) | 0.019(5) | 0.072(7) |
| | | u$_I$/a | 0.105(9) | 0.136(7) | 0.07(1) |
| | Global parameters: | E$_P$/k$_B$ (K) | 6.1(1) | 8.6(2) | 5.6(1) |
| | Statistics: | $\chi^2/\upsilon$ | 1.2714 | 6.0277 | 7.4397 |
| | $\upsilon$ = 62 | P (F-test) | – | 7.3×10$^{-9}$ | 8.6×10$^{-11}$ |
| (b) | Reference | Fitting parameter | Two-fluid | N(0)V = 4 | d-wave |
| | Powder A [4] | E$_P$/k$_B$ (K) | 0.9(8) | 0.90(1) | 1.46(2) |
| | H = 0.35 T | $\lambda(0)$ (nm) | 175(62) | 166.7(2) | 142.2(3) |
| | | u$_{p0}$/a | 0.07(18) | 0.094(0) | 0.146(1) |
| | | u$_{p1}$/a | 0.001(4) | 0.000(0) | 0.000(0) |
| | | u$_I$/a | 0.001(0) | 0.0015(0) | 0.001(0) |
| | | T$_C$ (K) | 90.63(6) | 94.16(7) | 94.53(8) |
| | Statistics: | $\upsilon$ | 11 | 13 | 13 |
| | | $\chi^2/\upsilon$ | 0.7289 | 17.6763 | 58.5557 |
| | | P (F-test) | – | 1.8×10$^{-5}$ | 5.5×10$^{-8}$ |
| | Powder B [48] | E$_P$/k$_B$ (K) | 17(3) | 4.7(9) | 0.4(1) |
| | H = 0.20 T | $\lambda(0)$ (nm) | 157(21) | 157.4(1) | 145.2(1) |
| | | u$_{p0}$/a | 0.05(10) | 0.003(1) | 0.056(0) |
| | | u$_{p1}$/a | 0.0013(6) | 0.000(0) | 0.000(0) |
| | | u$_I$/a | 0.048(4) | 0.033(7) | 0.000(0) |
| | | T$_C$ (K) | 89.46(9) | 92.89(9) | 95.0(1) |
| | Statistics: | $\upsilon$ | 6 | 7 | 7 |
| | | $\chi^2/\upsilon$ | 1.09323 | 25.1753 | 138.744 |
| | | P (F-test) | – | 0.0034 | 5.4×10$^{-5}$ |
| (c) | Universal parameter | Fitting parameter | Measurement A | Measurement B | |
| | Two-fluid | $\lambda_{ab}(0)$ (nm) | 133(40) | 133(40) | |
| | E$_P$/k$_B$ = 6.4(5) K | u$_{p0}$/a | 0.09(16) | 0.13(14) | |
| | $\chi^2/\upsilon$ = 1.7740 | u$_{p1}$/a | 0.167(4) | 0.188(2) | |
| | | u$_I$/a | 0.20(4) | 0.15(2) | |
| | N(0)V = 4 | $\lambda_{ab}(0)$ (nm) | 148.1(3) | 148.1(3) | |
| | E$_P$/k$_B$ = 8.1(3) K | u$_{p0}$/a | 0.000(0) | 0.090(3) | |
| | $\chi^2/\upsilon$ = 4.2760 | u$_{p1}$/a | 0.002(1) | 0.074(8) | |
| | P (F-test) = 0.019 | u$_I$/a | 0.253(3) | 0.195(3) | |
| | d-wave | $\lambda_{ab}(0)$ (nm) | 133.9(9) | 133.9(9) | |
| | E$_P$/k$_B$ = 5.0(1) K | u$_{p0}$/a | 0.047(9) | 0.110(5) | |
| | $\chi^2/\upsilon$ = 4.4044 | u$_{p1}$/a | 0.000(0) | 0.000(0) | |
| | P (F-test) = 0.015 | u$_I$/a | 0.206(6) | 0.163(3) | |

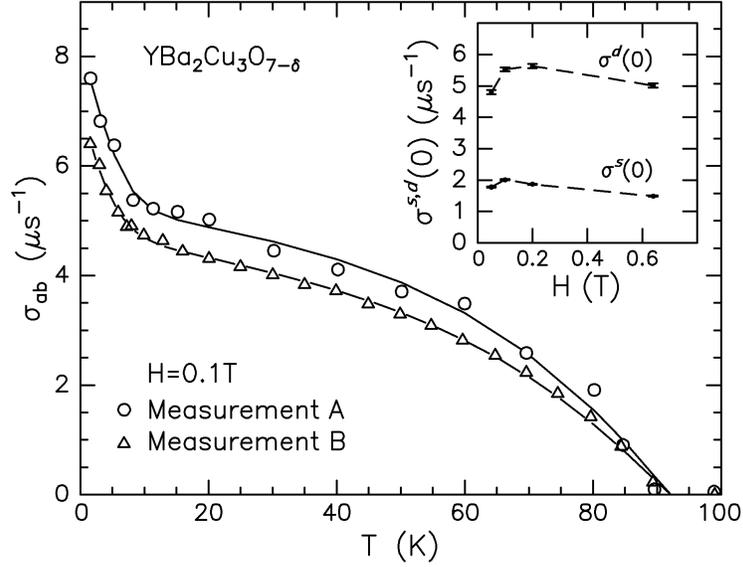

**Figure 3**. The *a-b* plane root second moment acquired on single-crystal YBa$_2$Cu$_3$O$_{7-\delta}$ sample. Measurement (A) is taken from figure 1 of [37]; Measurement (B) is taken from figure 3 of [37]. The two data sets were acquired on the same sample, at the same applied field (H=0.1 T) and over the same temperature range. Curves: fits of depinning model with a two-fluid gap function (parameters in table 1(c)). Inset: Variations of two-gap components $\sigma^d(0)$ and $\sigma^s(0)$ with external field H, from table I of [37] (notation "*s*" and "*d*" corrected).

In terms of the root second moment, a generalization of the two-fluid model is

$$\sigma(T) = \sigma(0)[1 - (T/T_C)^\alpha] . \qquad (12)$$

This functional form with $\alpha \approx 4$ has been shown to provide good fits to data for $T_C \approx$ 90-K phase of YBa$_2$Cu$_3$O$_{7-\delta}$ powders [3,4,49], which we interpret as corresponding to strong fluxon pinning, negligible influence of thermally-activated depinning, and an extremely strong-coupled *s*-wave pairing. Equation (12) assumes H << H$_{c2}$(T), which is applicable for $T_C - T > 0.2$ K. It is worth noting that this result is in excellent agreement with our analysis of single-crystal data in this paper and in previously reported work [25,26].

Additionally, inhomogeneous disorder produces an observable effect that is abundantly evident in oxygen-deficient YBa$_2$Cu$_3$O$_{7-\delta}$ (see figures 4 and 5 of [49] which show $\sigma(T)$ for oxygen content $0.003 \leq \delta \leq 0.484$ and transition temperatures 58.7 K $\leq T_C \leq$ 92.1 K). Fits using equation (12) to obtain the exponent $\alpha$ (see tables I and II in [49]) are plotted against $T_C$ in the inset of figure 2. The observed decrease in $\alpha$ (below the optimal value of $\alpha = 4$) as oxygen content is decreased and $T_C$ is depressed below 90K, and the sharp rise corresponding to the "60-K" oxygen-ordered phase, clearly shows that the exponent $\alpha$ is a good indicator of the effect of disorder on $\sigma(T)$. The resulting broadening of the observed transition temperature is accompanied by linear components in the temperature dependence in $\sigma(T)$. The increased $\Gamma$ associated with underdoping also correlates with weakened pinning seen, e.g. in the changed P(B) lineshape for low-temperature field-shifting in a twinned YBa$_2$Cu$_3$O$_{6.60}$ crystal (see figure 19 in [14]).

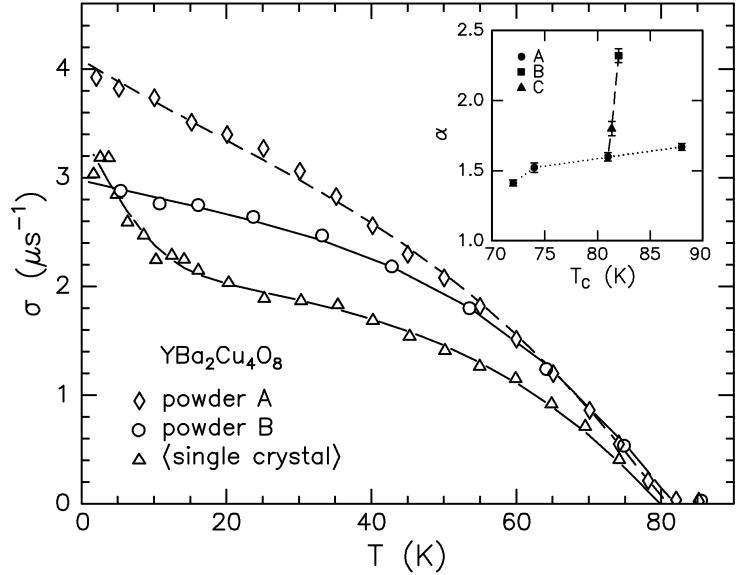

**Figure 4.** Root second moment data for $YBa_2Cu_4O_8$ specimens. The triangles represent the average of the field-oriented mosaic crystal data from [40] according to equation (11). The diamonds and circles correspond to the non-oriented powder data from (A) [52] (H = 0.2 Tesla) and (B) [49] (H = 0.35 Tesla, scaled by $2^{\frac{1}{2}}$, owing to the Gaussian form defined), respectively. Curves through data are fits of de-pinning model with two-fluid gap (parameters in table 2). Inset (powders): Fitted exponent α in equation (12) plotted against $T_C$; points A are for $YBa_2Cu_4O_8$ ($T_C$ = 81 K), $(Y_{0.94}Ca_{0.06})Ba_2Cu_4O_8$, ($T_C$ = 88 K), $Y(Ba_{1.925}La_{0.075})Cu_4O_8$ ($T_C$ = 74 K) and $Y(Ba_{1.9}La_{0.10})Cu_4O_8$ ($T_C$ = 72 K), extracted from [52] and connected by dotted lines; point B from [49]; point C from [54]. Dashed lines connect points for undoped powders.

*4.1.2. $YBa_2Cu_4O_8$* A similar $\mu^+$SR study was also conducted on an oriented mosaic of $YBa_2Cu_4O_8$ single crystals ($T_C$ = 79.9 K, $\Delta T_C \approx 2$ K, $\Gamma \sim 185$ [50]) for the magnetic field (H = 0.015 T) applied along each of the crystallographic axes, providing direct measurements of $\sigma_{ab}(T)$, $\sigma_{ac}(T)$ and $\sigma_{bc}(T)$ [40]. Since this material becomes optimal only under hydrostatic pressure ($T_{C0}$ = 104 K) [51][10], some perturbations due to non-optimization are expected. As in the case of single-crystal $YBa_2Cu_3O_{7-\delta}$, an inflection is observed in $\sigma_{ij}(T)$ at ~10 K, captured in all three planar components. Using equation (10) again, the axial components $\sigma_a(T)$, $\sigma_b(T)$ and $\sigma_c(T)$ were extracted [40], with the a and b components retaining the inflection while it is suppressed in the c component.

Similar to figure 2, we have reproduced in figure 4 the $\mu^+$SR data for two sintered powder samples of $YBa_2Cu_4O_8$: from (A) [52] ($T_C$ = 81 K, H = 0.2 T, open diamonds) and from (B) [49] ($T_C$ = 82.0(5) K, H = 0.35 T, open circles). These are compared to the average of the $\sigma_{ij}$ components for the oriented crystals (triangles). The random average of the oriented crystal data once again is observed to retain the low temperature inflection, whereas the sintered powder data show no such feature, but do show a clear departure from the two-fluid (α = 4) model. As in the case of $YBa_2Cu_3O_{7-\delta}$, this finding conforms to the first listed indicator of fluxon depinning effects, thus pointing to such phenomena as the origin of the inflection in σ(T). Moreover, the data for the two powder samples are not the same, reflecting the differing pinning structures contained within each. There is stronger departure from two-fluid behaviour than is observed for $YBa_2Cu_3O_{7-\delta}$. Since $\Gamma$ for this material is much greater than that for $YBa_2Cu_3O_{7-\delta}$, temperature-dependent fluxon depinning [53] and longitudinal disordering [35], even in powders, narrow the lineshape and contribute to the deviation from two-fluid ($\alpha \neq 4$).

---

[10] $T_{C0}$ extracted using midpoint of H = 0 resistance transition (figure 2 of [51]).

The data for figure 4 (three planar components for oriented crystal; two powders) have been fitted using our depinning model described above assuming the three possible underlying pairing states mentioned above. The resulting fitted parameters and statistics are given in table 2. The best fit of our depinning model for the crystalline sample data (component average shown by a solid curve in figure 4) assumes a two-fluid form for the gap function, as expressed in equation (8). Two-fluid model fits to the powder data are shown by dashed and solid curves for powders A [52] and B [49], respectively. The results of the F-distribution tests on the fits to all of the $YBa_2Cu_4O_8$ data show that the underlying gap model has not been determined with good confidence (probabilities associated with all lower-$\chi^2$ fits being greater than 0.04, which is non-negligible). For powder B data, the two-fluid or BCS models better explain the data, while the data for powder A with $\sigma(T \to 0)$ greater than that of both the randomly averaged oriented crystal or powder B data, are better fit with a *d*-wave model. It is likely that the stronger deviation $\sigma(T)$ from two-fluid and relatively strong variability of $\sigma(T,H)$ observed may be attributable to the fact that $YBa_2Cu_4O_8$ is optimized only under hydrostatic pressure. Due to the difficulties involved, $\mu^+SR$ experiments have not been conducted on the optimized compound, so no comparison is possible.

A glimpse into the effects of inhomogeneous disorder for the present case can be gleaned from figure 1 of [52], where the authors plot $\sigma(T)$ versus T for $YBa_2Cu_4O_8$ ($T_C$ = 81 K), $(Y_{0.94}Ca_{0.06})Ba_2Cu_4O_8$, ($T_C$ = 88 K), $Y(Ba_{1.925}La_{0.075})Cu_4O_8$ ($T_C$ = 74 K) and $YBa_{1.9}La_{0.10}Cu_4O_8$ $T_C$ = 72 K). Note that all of these compounds are non-optimal and have $T_C$'s far below the optimal $T_{C0}$ = 104 K [51]. By fitting these data to equation (12), one can plot the exponent $\alpha$ against $T_C$ as we have done in the inset in figure 4 (model function subtracts quadratically a fitted background $\sigma$ for each compound), where dotted lines connect points to guide the eye. It is clear that the shape of $\sigma(T)$ changes significantly with cation substitution, suggestive of the presence of disordering effects which force $\alpha$ lower as $T_C$ is suppressed below $T_{C0}$. Also plotted are exponents $\alpha$ fitted to published data for three undoped $YBa_2Cu_4O_8$ powder specimens, denoted as A [52] and B [49] (corresponding to the main figure) and as C [54][11]; these points are connected by the steep dashed lines to illustrate the trend. As one draws closer to the optimal superconducting state, thematerials become more homogeneous, $T_C$ and $\alpha$ increase, and the measured gap function reflects a more *s*-wave character. Non-optimization can be manifested in $\sigma(T,H)$, reflecting similar attributes as fluxon depinning and thus must be considered for non-optimal compounds.

A slight inflection observed in the temperature dependence of $\lambda_{ab}^{-2}(T)$ derived from low-field magnetization of a *c*-axis-aligned polycrystalline $YBa_2Cu_4O_8$ sample [55] (singled out as an example of "powder samples" in [40]). As the authors observe, this anomalous behaviour is not endemic to high-$T_C$ compounds and is likely to be materials related [55]. Indeed, the sample preparation described all but guarantees a sample filled with dislocations and cracks; flux penetration into the cracks and the formation of Josephson vortices could mimic pinning effects of the mixed state even for $H < H_{c1}$, easily explaining the anomalous magnetization results. Such a sample is obviously not a random sintered powder [49,52], and measurements of diamagnetic screening at surfaces are not equivalent to the bulk probe provided by $\mu^+SR$ [40]. Such presumed equivalences in sample morphologies as well as measured physical properties serve only to obviate the approach of [40] to meaningful data comparison.

---

[11] $T_C$ = 81.4(5) K from fitting to equation (12).

**Table 2.** (a) Depinning model fit results for single-crystal $YBa_2Cu_4O_8$ ($T_C$ fixed at 79.9 K). Data from [40]. F-test probabilities are relative to best-fitting model. (b) Depinning model fit results for powder samples of $YBa_2Cu_4O_8$.

| (a) | Component | Fitting parameter | Two-fluid | $N(0)V = 4$ | $d$-wave |
|---|---|---|---|---|---|
| | $ab$-plane | $\lambda_{ab}(0)$ (nm) | 123(9) | 180.2(3) | 134(1) |
| | H = 0.015 T | $u_{p0}/a$ | 0.22(2) | 0.000(0) | 0.169(5) |
| | | $u_{p1}/a$ | 0.182(5) | 0.000(0) | 0.008(1) |
| | | $u_I/a$ | 0.24(2) | 0.51(1) | 0.254(4) |
| | $bc$-plane | $\lambda_{bc}(0)$ (nm) | 391(8) | 378(94) | 361(4) |
| | H = 0.015 T | $u_{p0}/a$ | 0.05(2) | 0.09(14) | 0.01(3) |
| | | $u_{p1}/a$ | 0.10(3) | 0.005(1) | 0.006(1) |
| | | $u_I/a$ | 0.17(1) | 0.16(4) | 0.05(1) |
| | $ac$-plane | $\lambda_{ac}(0)$ (nm) | 409(12) | 205(41) | 420(4) |
| | H = 0.015 T | $u_{p0}/a$ | 0.12(2) | 0.35(4) | 0.005(17) |
| | | $u_{p1}/a$ | 0.10(3) | 0.000(0) | 0.011(2) |
| | | $u_I/a$ | 0.07(1) | 0.028(6) | 0.000(0) |
| | Global Parameters: | $E_P/k_B$ (K) | 9.3(4) | 9.4(2) | 7.2(2) |
| | Statistics: | $\chi^2/\upsilon$ | 1.3251 | 1.5596 | 2.23803 |
| | $\upsilon = 54$ | P (F-test) | – | 0.56 | 0.059 |
| (b) | Sample | Fitting parameter | Two-fluid | $N(0)V = 4$ | $d$-wave |
| | Powder A [52] | $E_P/k_B$ (K) | 12.8(1) | 43(2) | 19(1) |
| | | $\lambda(0)$ (nm) | 159.0(2) | 186(1) | 159.2(2) |
| | H = 0.2 T | $u_{p0}/a$ | 0.002(0) | 0.10(1) | 0.021(0) |
| | $T_C$ = 81 K (fixed) | $u_{p1}/a$ | 0.244(1) | 0.000(0) | 0.121(2) |
| | | $u_I/a$ | 0.000(0) | 0.147(1) | 0.000(0) |
| | Statistics: | $\upsilon$ | 12 | 12 | 12 |
| | | $\chi^2/\upsilon$ | 2.3302 | 1.3109 | 0.64825 |
| | | P (F-test) | 0.044 | 0.26 | – |
| | Powder B [49] | $E_P/k_B$ (K) | 13.6(1) | 23(2) | 11(1) |
| | | $\lambda(0)$ (nm) | 184.9(4) | 198.2(5) | 177.3(2) |
| | H = 0.35 T | $u_{p0}/a$ | 0.002(0) | 0.000(0) | 0.062(1) |
| | $T_C$ = 82 K (fixed) | $u_{p1}/a$ | 0.180(3) | 0.000(0) | 0.000(0) |
| | | $u_I/a$ | 0.000(0) | 0.068(3) | 0.000(0) |
| | Statistics: | $\upsilon$ | 4 | 4 | 4 |
| | | $\chi^2/\upsilon$ | 1.4394 | 1.0179 | 9.4625 |
| | | P (F-test) | 0.78 | – | 0.10 |

*4.1.3. $La_{1.83}Sr_{0.17}CuO_4$* The $\mu^+$SR study for the $La_{1.83}Sr_{0.17}CuO_4$ crystal ($T_C$ = 36.2 K, $\Delta T_C$ = 1.5 K) comprises measurements of $\sigma_{ab}(T,H)$ as a function of both temperature and magnetic field (*ac*- and *bc*-components not given) [41]. The crystal is slightly overdoped (optimal stoichiometry is $La_{1.837}Sr_{0.163}CuO_4$ for which $T_{C0}$ = 38.5 K [56] and $\Gamma \approx 170$ [57]). Measurements of the temperature variation of $\sigma_{ab}(T,H)$ for applied magnetic fields of 0.02, 0.1 and 0.64 T show a clear inflection in $\sigma_{ab}(T)$ occurring at ~10 K at 0.02 T, which is completely suppressed at 0.64 T [data from figure 2 of [41] is reproduced in figure 5(a)]. Meanwhile $\sigma_{ab}(T\rightarrow 0,H)$ decreases by about 37% for a field change from 0.02 to 0.64 T (see figure 6, triangles). For comparison, $\sigma(T)$ data acquired on a powder sample of $La_{1.85}Sr_{0.15}CuO_4$ at applied fields 0.1 and 0.2 T [58] are shown in figure 5(b). The obvious differences observed for $\sigma(T)$ between the powder and single-crystal samples indicate that, as was found for $YBa_2Cu_3O_{7-\delta}$ and $YBa_2Cu_4O_8$, the anomalous inflection at low temperatures cannot be intrinsic to the gap function, but is attributable to fluxon depinning (first indicator listed in section 1).

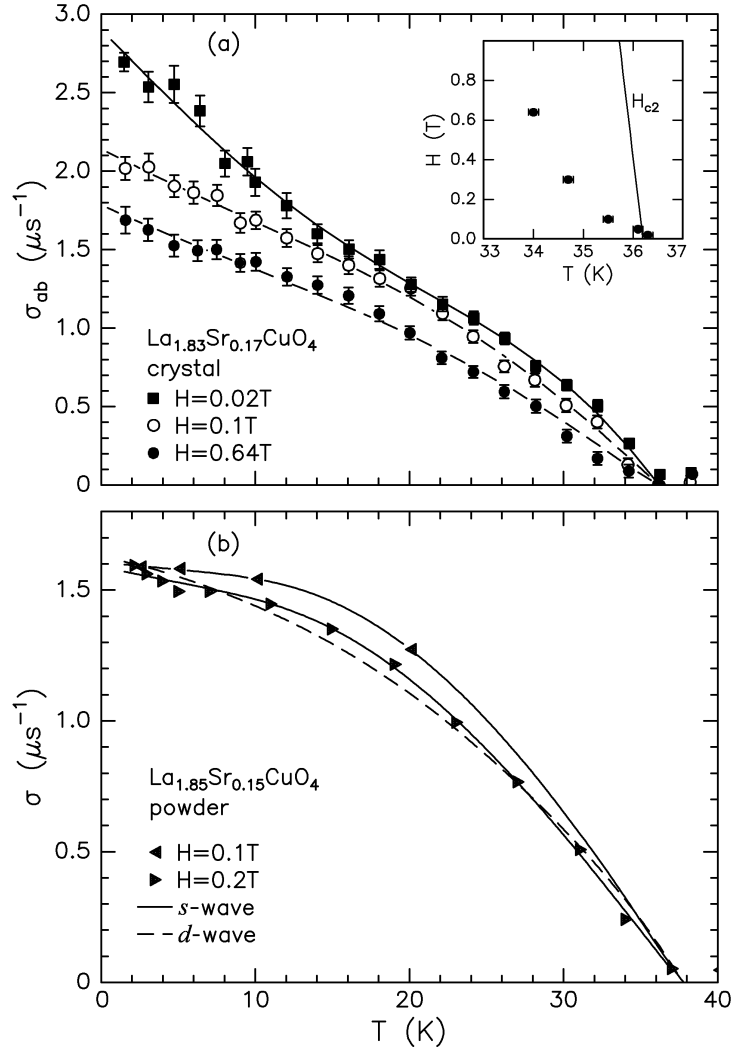

**Figure 5.** Frame (a) $\sigma_{ab}$ versus temperature for $La_{1.83}Sr_{0.17}CuO_4$ crystal measured at three fields (data points from [41]); the curves represent the depinning model assuming a two-fluid gap (parameters in table 3(a)). Frame (b) shows comparative data for a powder sample of $La_{1.85}Sr_{0.15}CuO_4$ at applied fields of 0.1 and 0.2 T taken from [58]. The curves through the data correspond to the depinning model assuming BCS gap with $N(0)V = 4$ (*s*-wave, solid) and a *d*-wave gap (dashed) function (fitted parameters for two-fluid, $N(0)V = 4$ and a *d*-wave gap are given in table 3(b)). Inset in frame (a): (points) "$T_C$" from two-gap model, table I of [41]; (line) upper critical field $H_{C2}(T)$ (slope from [59-61]).

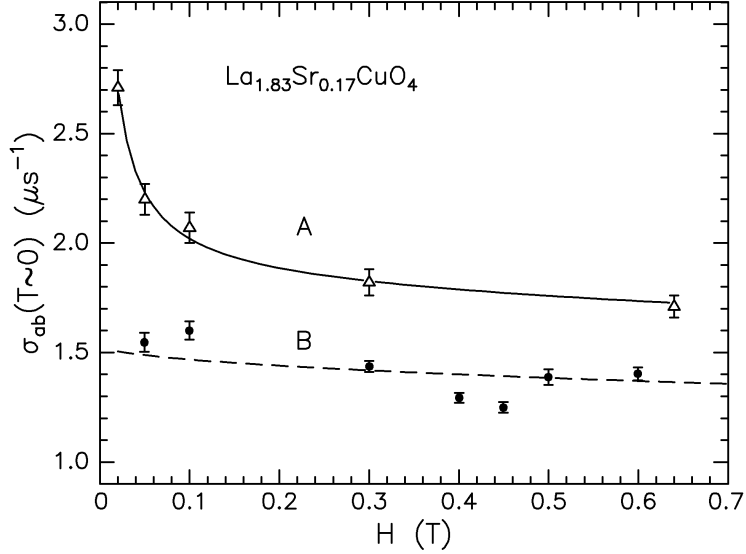

**Figure 6.** Plot of $\sigma_{ab}(T\sim 0)$ versus applied field H for $La_{1.83}Sr_{0.17}CuO_4$. Data set "A" is taken from [41] and data set "B" is from [63]. The solid curve is the depinning model fitted to data "A"; dashed curve is for ideal flux lattice [44] (absent depinning effects) fitted to data "B".

The solid and dashed curves in figure 5(a) represent the fit of our depinning model to the single-crystal data assuming the two-fluid model. The resulting parameters for the two-fluid, BCS with $N(0)V = 4$ and $d$-wave pairing models are given in table 3(a). The F-distribution test shows that comparably good fits are obtained with the three pairing models, which indicates that pinning perturbations in this experiment prevent us from discerning the underlying gap symmetry. For the three fields shown, the fluxon lattice spacings, relative to the London penetration depth, are $a/\lambda_L = 1.57, 0.703$, and $0.278$ for $H = 0.02, 0.1$ and $0.64$ T, respectively (assuming $\langle B \rangle \approx H$). Note that only for $H = 0.02$ T is the fluxon lattice spacing greater than $\lambda_L$, which would enhance the depinning effects compared to higher field.

The two-gap model used in [41] to fit the same data yields dependence of an apparent "$T_C$" on applied field (by way of contrast, our depinning model is based on a single true $T_C = 36.2$ K, as determined by magnetization [41]). This is shown by the inset to figure 5(a), where the ordinate H is the applied field and the abscissa for the points are the "$T_C$" values given in table I of [41]. The line shows the $H_{C2}(T)$ phase boundary, which is drawn through the true $T_C$ with slope 2.1 T/K, as given by the mean of three independent measurements ($dH_{c2}/dT|_{Tc} = 2.13$ [59], 2.1 [60] and 2.2 T/K [61]). The large discrepancy between the ("$T_C$",H) data points and $H_{c2}(T)$ illustrates the fourth indicator of fluxon depinning effects presented in section 1.

Similarly, the solid curves shown in figure 5(b) represent the best fit of our depinning model to the powder data of [58] for the two fields studied, which is the BCS form with $N(0)V = 4$. Fitted parameters and statistics are given in table 3(b). F-distribution tests show that BCS and two-fluid models give comparably good fits, whereas the $d$-wave model has low probability (0.0045) of giving a good fit. The fit obtained for 0.2 T with the $d$-wave gap function is given by the dashed curve in figure 5(b), illustrating its deviation from the data.

While prior work concerning fluxon depinning effects in single-crystal $YBa_2Cu_3O_{7-\delta}$ [25] was acknowledged, pinning was nevertheless treated insufficiently in [41] with a Gaussian convolution model, which merely corrects for minor inhomogeneities in sample magnetization. This approach was rationalized by asserting that pinning leads to an almost symmetric lineshape and noting that the observed lineshapes are asymmetric (figure 1 of [41]). However, as noted in [14] such asymmetry in P(B) is consistent with pinning. In actuality, the deviations observed [41] in the high-field tail of the local magnetic field distribution are most certainly due to field-induced changes in the pinning structure, and a symmetric lineshape is only observed with vortex melting at T ≈ 30 K (the single-crystal $La_{1.83}Sr_{0.17}CuO_4$ sample studied in [41] was the same sample measured earlier by different groups [62,63]). Even for static pinning, lineshapes depend upon the structure of the pinning potential [2], crystal stacking faults and the longitudinal flux line correlation length [33,35]; a symmetric and greatly narrowed lineshape is, for example, observed for $Bi_2Sr_2CaCu_2O_{8+\delta}$ at 1.5 T (see figure 3 of [35]), and is attributable to a shortened hard-axis (longitudinal) correlation length.

**Table 3.** (a) Depinning model fit results for single-crystal $La_{1.83}Sr_{0.17}CuO_4$ ($T_C$ fixed at 36.2 K). Data from [41]. F-test probabilities are relative to the best-fitting model. (b) Depinning model fit results for powder sample of $La_{1.85}Sr_{0.15}CuO_4$. Data taken from [58].

| (a) | Model | Fitting parameter | H=0.02T | H=0.1T | H=0.64T |
|---|---|---|---|---|---|
| | Two-fluid | $E_P/k_B$ (K) | 14.3(5) | 14.3(5) | 14.3(5) |
| | $\upsilon = 47$ | $\lambda_{ab}(0)$ (nm) | 220(1) | 220(1) | 220(1) |
| | $\chi^2/\upsilon = 0.73366$ | $u_{p0}/a$ | 0.212(3) | 0.016(4) | 0.098(5) |
| | | $u_{p1}/a$ | 0.004(1) | 0.255(2) | 0.262(3) |
| | | $u_l/a$ | 0.256(4) | 0.001(0) | 0.002(0) |
| | N(0)V = 4 | $E_P/k_B$ (K) | 10.8(4) | 10.8(4) | 10.8(4) |
| | $\upsilon = 47$ | $\lambda_{ab}(0)$ (nm) | 241(3) | 241(3) | 241(3) |
| | $\chi^2/\upsilon = 0.89877$ | $u_{p0}/a$ | 0.116(8) | 0.05(1) | 0.015(15) |
| | P (F-test) = 0.49 | $u_{p1}/a$ | 0.003(3) | 0.15(1) | 0.217(6) |
| | | $u_l/a$ | 0.294(7) | 0.10(1) | 0.03(1) |
| | $d$-wave | $E_P/k_B$ (K) | 10.9(5) | 10.9(5) | 10.9(5) |
| | $\upsilon = 47$ | $\lambda_{ab}(0)$ (nm) | 215(2) | 215(2) | 215(2) |
| | $\chi^2/\upsilon = 0.86611$ | $u_{p0}/a$ | 0.100(6) | 0.091(8) | 0.126(6) |
| | P (F-test) = 0.58 | $u_{p1}/a$ | 0.003(1) | 0.113(9) | 0.167(5) |
| | | $u_l/a$ | 0.197(3) | 0.039(20) | 0.001(0) |
| (b) | Model | Fitting parameter | H=0.1T | H=0.2T | |
| | Two-fluid | $E_P/k_B$ (K) | 1.2(1) | 1.2(1) | |
| | $\upsilon = 7$ | $\lambda(0)$ (nm) | 250.6(5) | 250.6(5) | |
| | $T_C = 36.3(1)$ K | $u_{p0}/a$ | 0.029(8) | 0.014(3) | |
| | $\chi^2/\upsilon = 1.4428$ | $u_{p1}/a$ | 0.135(7) | 0.177(3) | |
| | P (F-test) = 0.31 | $u_l/a$ | 0.001(0) | 0.001(0) | |
| | N(0)V = 4 | $E_P/k_B$ (K) | 1.0(1) | 1.0(1) | |
| | $\upsilon = 8$ | $\lambda(0)$ (nm) | 250.5(4) | 250.5(4) | |
| | $T_C = 37.8(1)$ K | $u_{p0}/a$ | 0.047(7) | 0.042(2) | |
| | $\chi^2/\upsilon = 0.64078$ | $u_{p1}/a$ | 0.10(1) | 0.147(4) | |
| | | $u_l/a$ | 0.000(0) | 0.001(0) | |
| | $d$-wave | $E_P/k_B$ (K) | 0.4(1) | 0.4(1) | |
| | $\upsilon = 8$ | $\lambda(0)$ (nm) | 242.9(3) | 242.9(3) | |
| | $T_C = 37.73$K | $u_{p0}/a$ | 0.046(1) | 0.069(2) | |
| | $\chi^2/\upsilon = 7.4089$ | $u_{p1}/a$ | 0.000(0) | 0.000(0) | |
| | P (F-test) = 0.0045 | $u_l/a$ | 0.0014(0) | 0.0061(0) | |

Local magnetic field distributions P(B) for H = 0.05 and 0.64 T given in figure 1 of [41] show that the internal fields B corresponding to the peak and low-field cut-off in P(B) do not change with H (at T=1.7 K). From the theoretical model for P(B) used in [41] (see also figure 11 in [44]) one finds that the average internal field ⟨B⟩ is about 0.1 mT greater than the external field H; thus equilibrium diamagnetism of the mixed state is not observed. A similar result, ⟨B⟩ ≥ H at low temperature, was reported in [28]; there it was identified as a symptom of fluxon pinning. Furthermore, the posited field-induced transition from a triangular to a square lattice discussed in [41] and [63] would shift the peak in P(B) by +0.33 mT and the cut-off by −0.57 mT for ideal fluxon lattices, whereas neither effect is observed to within the experimental resolution, ∼ 0.15 mT. Distortion of the fluxon lattice by pinning, if sufficiently pronounced, could obscure lattice symmetry features in P(B) in addition to modifying the diamagnetism.

The fully static pinning model of [41] does not take into account temperature-activated fluxon depinning and is of insufficient generality to explain the non-monotonic variation of $\sigma_{ab}(T{\sim}0,H)$ observed in $La_{1.83}Sr_{0.17}CuO_4$ (e.g. a minor maximum at 0.1 T and a minimum at 0.45 T) [63] and $YBa_2Cu_3O_{7-\delta}$ [25,37]. This non-monotonic response is illustrated for $La_{1.83}Sr_{0.17}CuO_4$ in figure 6, where data obtained from [63] (reproduced as filled circles) show a markedly subdued and non-monotonic field dependence, when compared to the lesser complete data of [41] (triangles). The solid curve in figure 6 shows the depinning model for $\sigma_{ab}(T=0,H)$ at H = 0.02, 0.1, and 0.64 T and smooth interpolation for fields in between points. The dashed curve in figure 6 is the intrinsic function for $\sigma_{ab}(T=0,H)$ obtained by fitting equations (7)–(9) to data from [63] for H ≤ 0.3 T, with $H_{c2}(0)$ = 82 T and $\sigma_{ab}(T=0,H=0)$ = 1.60 ± 0.4 $\mu s^{-1}$. The apparent irreproducibility exhibited by the two sets of data points in figure 6, which were obtained at the same experimental facility for ostensibly the same crystal, can be understood as differing manifestations of pinning in the two experiments (second indicator listed in section 1).

*4.2. $YBa_2Cu_3O_{7-\delta}$ and $YBa_2Cu_4O_8$: the c-axis component*

Accepting, as one now must, that the inflection in σ(T) observed at ∼10–20 K in the high-$T_C$ compounds is due to temperature-activated fluxon depinning, one would then expect the obscuring effects of fluxon instabilities to be anisotropic, since fluxon motion is restricted along the hard axis. To understand the implication, we decompose the pinning factor in equation (1) into axial components as $f_p = (f_if_j)^{1/2}$ for i ≠ j = a, b, or c, where $f_a$ and $f_b$ are temperature dependent. The component $f_c$ is taken to be constant, owing to strong intrinsic pinning for fluxon axes parallel to the *ab* plane [64], restricting vortex movement along the *c*-axis. Thus in determining the *c*-axis component $\sigma_c$ according to equation (10), the *ab*-planar pinning components $f_a$ and $f_b$ cancel out, so that $f_p = f_c$ and remains constant with temperature. Consequently, the *c*–axis component $\sigma_c(T)$ would best reflect the true (underlying) gap function.

The temperature dependent *c*–axis components for both $YBa_2Cu_3O_{7-\delta}$ [37] and $YBa_2Cu_4O_8$ [40] are given in figure 7; the curves shown are fits of the generalized two-fluid model [equation (12)], yielding α = 3.30(17) and $\sigma_c(T{\to}0)$ = 0.1845(22) $\mu s^{-1}$ for $YBa_2Cu_3O_{7-\delta}$, and α = 4.5(3) and $\sigma_c(T{\to}0)$ = 0.0965 $\mu s^{-1}$ for $YBa_2Cu_4O_8$ (the latter values taken from [40]). Since the $YBa_2Cu_4O_8$ sample is non-optimal, the significance of the actual value of α is unclear. Nevertheless, taking a weighted average of the two exponents above yields ⟨α⟩ = 3.6 ± 0.6, consistent with the classical two-fluid value (α = 4). We point out that the scatter below ∼10 K for the $YBa_2Cu_3O_{7-\delta}$ is likely due to the mismatch between the fields applied perpendicular and parallel to the *a-b* plane. Unlike the disordering exhibited by the cation-substituted high-$T_C$ compounds, the fact that $\sigma_c(T)$ for $YBa_2Cu_4O_8$ (at ambient pressure) exhibits an α exponent consistent with two-fluid behaviour indicates that the effects of non-optimization in this material (due to an absence of the necessary hydrostatic pressure) are homogeneously distributed with respect to the superconductivity.

**5. Discussion**

Justification for the above fluxon pinning analysis of the σ(T,H) data for three crystal and six sintered-powder specimen types, and the applicability of the thermal depinning model of [25,26] for determining the superconducting penetration depth, can be found in scanning tunnelling microscopy (STM) studies of superconducting *a-b* surfaces in the mixed state (e.g. of $YBa_2Cu_3O_{7-\delta}$ crystals) [24]. These microscopic studies reveal disordered inter-vortex spacings, which are represented in the depinning model by $u_l$, and apparent enlargement of vortex cores, represented by $u_p$ [see equations (2-3)]. However, in recognition of the extrinsic nature and materials dependence of pinning [65], one might consider whether the depinning model is of sufficient generality, even though it follows from well-developed theory of fluxon properties [2]. We address this issue by noting consistencies in extracted penetration depths and mass anisotropy ratios, which provide quantitative checks confirming the model's validity. In all cases where statistically significant fits are obtained, the T-dependence of the penetration depth is consistent with a nodeless gap function (corresponding to the superconducting hole condensate). Of the 45 analytical tests performed for this study a best fit with *d*-wave pairing symmetry is obtained for only one sample (non-optimal $YBa_2Cu_4O_8$ powder A [52]) and in that case with low statistical significance [table 2(b)]. The following draws the inferences from the results and considers the implications for pairing symmetry concepts.

*5.1. Fluxon pinning and thermal depinning*

Seemingly broad variation of pinning effects among various compounds and different sample types depends on two factors: (1) the distribution of crystalline defects and (2) the mass anisotropy ratio $\Gamma$. Typically, if the defect density is large, one would likely observe strong pinning, in which case $\sigma(T,H)$ would better reflect the true gap function as given by the root second moment $\sigma_0(T,H)$ intrinsic to the fluxon lattice. Powder samples of $YBa_2Cu_3O_{7-\delta}$ and the other cuprates generally fall into this category. If, on the other hand, the defect density is low, the fluxon lattice would likely experience very weak pinning. Materials falling into this limiting regime are single-crystal samples of $YBa_2Cu_3O_{7-\delta}$, $YBa_2Cu_4O_8$, $La_{2-x}Sr_xCuO_4$, $Bi_2Sr_2CaCu_2O_{8+\delta}$ [28] and $\kappa$–[BEDT-TTF]$_2$Cu[NCS]$_2$ [66,67] ($T_C$ = 10.5 K, $\Gamma = m^*_a/m^*_{bc} > 4\times10^4$ [68]), in order of increasing $\Gamma$. The temperature dependence of $\sigma(T,H)$ is typically characterized by $\alpha < 4$ in equation (12), and a strongly field-dependent (and possibly non-monotonic) $\sigma(T\rightarrow 0,H)$, obscuring the true character of the gap symmetry.

Differences between single-crystal and powder data sets shown in figures 2, 4, and 5 are readily understood as arising from differing defect densities and pinning energy distributions. Apart from the inflection at ~10 K for $YBa_2Cu_3O_{7-\delta}$ and $YBa_2Cu_4O_8$ being present only in crystal data, figures 2 and 4 also reveal that the σ(T→0) values acquired on the powder samples differ significantly from that of the randomly averaged crystalline data, which are both strongly field dependent. The influence of pinning on σ(T→0) was recognized in $YBa_2Cu_3O_{7-\delta}$ crystals in [25] by the observation of a strong non-monotonic field dependence; it is also seen in [37], as exemplified by the inset to figure 3, and in $YBa_2Cu_4O_8$ [40].

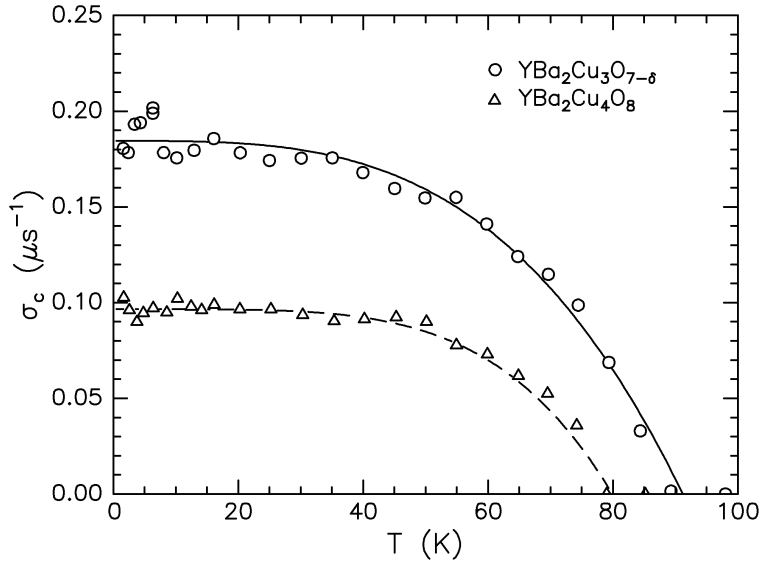

**Figure 7**. The extracted *c*-axis relaxation component $\sigma_c(T)$ acquired on single-crystal $YBa_2Cu_3O_{7-\delta}$ sample (circles), taken from [37], and oriented mosaic crystal $YBa_2Cu_4O_8$, taken from [40] (triangles). The curves through the data represent the best fit of the generalized two-fluid model equation (12); for $YBa_2Cu_3O_{7-\delta}$, $\sigma(T\rightarrow 0) = 0.1845(22)$ $\mu s^{-1}$ and $\alpha = 3.30(17)$ and for $YBa_2Cu_4O_8$, $\sigma(T\rightarrow 0) = 0.0965$ $\mu s^{-1}$ and $\alpha = 4.5(3)$ (latter fitted values from [37]).

For the $La_{1.83}Sr_{0.17}CuO_4$ crystal data set B in figure 6 (filled circles) also show non-monotonicity of $\sigma_{ab}$(T→0,H), which is observed to fluctuate about the theoretical dashed curve (intrinsic field dependences for alternative pairing and lattice symmetries have similar forms, see [44]). Fluxon pinning, seen as much larger $\sigma_{ab}$(T→0,H) in data set A, is a plausible explanation for the strong 37% variation of $\sigma_{ab}$(T→0,H) with applied field, when compared to the much weaker 7.6% theoretical intrinsic variation. Moreover, the result for $\sigma_{ab}$(T→0, H=0.02 T) in data set A is significantly larger (about a factor of 1.7) than the extrapolated trend of the data in set B. This is directly displayed in plots of P(B) in [41], which show it to be broader than in [63]. These signatures of data irreproducibility, which are unlikely to arise solely from data reduction, show that P(B) is non-unique, possibly because of variable field-cooling procedures. Enhancements in $\sigma_{ab}$, such as exhibited by data set A, produce erroneous estimates of the penetration depth when pinning is not taken into account [13]. One may also note that the unshifted P(B) and ⟨B⟩ observed for field-shifting studies in [14,36] are consistent with the presence of a thermal depinning component in fitted $\lambda_{ab}$(T) (YBCO1 in figure 1) (i.e. short-range thermal reordering of fluxons at nearly constant average flux density).

The similarities of the low-temperature inflection features observed for $YBa_2Cu_3O_{7-\delta}$, $YBa_2Cu_4O_8$, $La_{1.83}Sr_{0.17}CuO_4$ and $Bi_2Sr_2CaCu_2O_{8+\delta}$ crystals can be understood from the depinning model: The pinning energies $E_p$ that characterize smooth fluxon line displacements are comparable among these cuprates (cuprates may contain similar defect types and have comparably very small coherence distances) and vary weakly with Γ as expected; only when Γ becomes large does longitudinal disorder become important, inducing a truncated c-axis correlation length and a narrowing of σ(T,H) [28] with increasing H. While the minimum value of Γ required for fluxon depinning to occur in the high-$T_C$ compounds is not precisely known, it is certainly less than ~25 [5] (as such phenomena are observed in $YBa_2Cu_3O_{7-\delta}$).

It is also worth noting that the κ–[BEDT-TTF]$_2$Cu[NCS]$_2$ data from [66] were acquired in a transverse field of 129 Oe, and exhibit a $T_C$ consistent with 10.5 K, whereas data taken earlier [67] at a higher field of 0.3 T present a depressed $T_C$, reflecting a trend consistent with fluxon dynamics in the presence of high Γ. More importantly, figure 4(b) of [66] shows a minimum in ⟨ΔB(T)⟩ at ~5 K, in direct correspondence to the associated anomaly in ⟨|ΔB(T)|$^2$⟩, and similar to that observed for $Bi_2Sr_2CaCu_2O_{8+\delta}$ [28].

For the $La_{1.83}Sr_{0.17}CuO_4$ crystal the anomalously large field dependence in apparent "$T_C$", shown in the inset of figure 5(a), was obtained in [41] by fitting $\sigma_{ab}$(T,H) over the full temperature range in the data. Consequently, rather than an actual depression in $T_C$, a signature of temperature-dependent pinning has been presented.

*5.2. Penetration depth analysis*

Consistency in London penetration depths (i.e. $\lambda_{ab}$ in the limits T→0, H→0) obtained by applying the depinning model to data for crystals and powders confirms that valid results are determined from the second moment for both types of specimens. For $YBa_2Cu_3O_{7-\delta}$, a good (modern) powder value of λ(0) = 157(21) nm is given for powder B in table 1(b), which corresponds (according to [13]) to an a-b plane value of $\lambda_{ab}$(0) = 127(17) nm, in agreement (within < 5%) of the single-crystal value $\lambda_{ab}$(0) = 133(40) nm from table 1(c). Also, from statistically weighted averages the results in table 3(a) give $\lambda_{ab}$(0) = 224(11) nm for the $La_{1.83}Sr_{0.17}CuO_4$ single crystal and table 3(b) gives λ(0) = 251(1) nm for the $La_{1.85}Sr_{0.15}CuO_4$ powder; the ratio is 0.89(4), which compares favourably with the theoretical result of 0.81 calculated in [13] (uncorrected for slightly different Sr contents and anisotropic diamagnetism).

Our depinning model analysis also yields consistent results for the mass anisotropy $\Gamma = m^*_c/m^*_{ab}$. This is obtained as $[\lambda_c(0)/\lambda_{ab}(0)]^2 = \sigma_{ab}(T\rightarrow 0)/\sigma_c(T\rightarrow 0)$ by orienting the transverse magnetic field away from the hard axis of anisotropic superconductors (given static fluxons at low temperature). Such an experiment was first performed on twinned single-crystals of $YBa_2Cu_3O_{7-\delta}$ with the field angled at 45 degrees from the *c*-axis, resulting in $\Gamma \geq 25$ [5]. From figure 3 of [37], the $\gamma_{ca}$ and $\gamma_{cb}$ anisotropies (defined therein as ratios of $\sigma^{1/2}$) at lowest temperatures have a geometrical average $(\gamma_{ca} \gamma_{cb})^{1/2} \approx 5.8$, corresponding to an average mass anisotropy $\Gamma \approx 33.6$. Moreover, by calculating [using equation (7)] the *c*-axis component penetration depth corresponding to $\sigma_c(T\rightarrow 0) = 0.183$ $\mu s^{-1}$, one obtains $\lambda_c(0) \approx 766.2$ nm, which, when compared to the ab-plane value of $\lambda_{ab}(0) = 133.0$ nm obtained from table 1(c), yields $\Gamma \approx 33.2$ (furthermore $\lambda_{ab}(0)$ agrees well with the result $127.6 \pm 1.5$ nm obtained in [25] for a similar $YBa_2Cu_3O_{7-\delta}$ crystal at four fields). These rather remarkable agreements support the accuracy of our depinning-model interpretation. While the temperature and field (strength and orientation) dependences of fluxon pinning are exceedingly important in understanding and interpreting the $\mu^+SR$ data, the $\gamma$-anisotropies in $\sigma$ are dominated by the superconducting mass anisotropy $\Gamma$ and rather insensitive to pinning differences.

Another important validation of our depinning model, and the observation that the intrinsic pinning associated with $\sigma_c(T)$ is temperature independent, is illustrated for $YBa_2Cu_3O_{7-\delta}$ by the excellent agreement between $\lambda_c(0) = 766.2$ nm, calculated from $\sigma_c(T\rightarrow 0)$ (uncorrected for pinning), and $\lambda_c(0) = [\lambda_{ac}(0) \lambda_{bc}(0)]/\lambda_{ab}(0) = 774.7$ nm, from tables 1(a) and 1(c). We use the value of $\lambda_{ab}(0)$ from table 1(c) since it represents a measurement at one unique field (0.1 T), not a result of analyzing mixed-H data.

*5.3. Dominant superconducting gap symmetry*

Application of the depinning model to the more recent $YBa_2Cu_3O_{7-\delta}$ crystal data [37] provides a best fit with a single superconducting gap modelled by the two-fluid function, indicative of strong-coupled nodeless (consistent with *s*-wave) superconductivity, in complete agreement with the earlier crystal results [25]. Thus, in studies of two single-crystal $YBa_2Cu_3O_{7-\delta}$ samples the F-distribution tests (table 1(a), figure 5 in [25]) confirm negligibly small probabilities that *d*-wave theory fits the $\mu^+SR$ data. Analyses of the $YBa_2Cu_3O_{7-\delta}$ powder data yield the same conclusion [table 1(b)]; the plainly evident nodeless behaviour is fully substantiated, since depinning perturbations are either very small or practically absent. The two-fluid form of $\sigma_c(T)$ provides further evidence for nodeless gap symmetry in $YBa_2Cu_3O_{7-\delta}$.

Applying a similar pinning analysis of the ambient-pressure (i.e. non-optimal) $YBa_2Cu_4O_8$ oriented crystal mosaic and powder data shows that they are best described by the same fluxon depinning model, even though statistical analysis may not satisfactorily distinguish the gap symmetry (among the models tested, the probability of *d*-wave is only ~6%) from the temperature dependence of the penetration depth. A similar finding is obtained for the slightly non-optimal $La_{1.83}Sr_{0.17}CuO_4$ crystal, although a nodeless symmetry is readily distinguished in the powder data.

Nodeless gap symmetry was previously shown to be on full display in the penetration depth obtained directly from $\mu^+SR$ data for the high-$T_C$ organic superconductor, $\kappa-[BEDT-TTF]_2Cu[NCS]_2$, where $\lambda_{bc}(T)$ is constant to a precision of 3% at temperatures below 2.5 K; this is consistent with *s*-wave pairing and rules out a *d*-wave gap [66]. In this case, fluxon depinning occurs at $T > T_C/2$, (instead of $T > T_C/6$ for $YBa_2Cu_3O_{7-\delta}$) allowing for a wide enough temperature range below the depinning temperature for the true gap-driven temperature dependence in $\lambda_{bc}(T)$ to be revealed.

Other evidence for *s*-wave pairing, as well as certain inconsistencies with *d*-wave symmetry, has been found in scanning tunnelling microscopy studies of *a-b* plane surfaces of $Bi_2Sr_2CaCu_2O_{8+\delta}$ and $YBa_2Cu_3O_{7-\delta}$. Hasegawa et al. have observed very flat-bottomed tunnelling spectra with differential conductance near 1% at zero-bias and well-defined (although broadened) peaks at ± voltage biases corresponding to the superconducting energy gap ($Bi_2Sr_2CaCu_2O_{8+\delta}$ crystals, $YBa_2Cu_3O_{7-\delta}$ epitaxial films), which are the hallmark features of an *s*-wave gap [69,70].[12] In applied fields other studies have shown sub-gap peaks at ~5.5 meV for tunnelling into the vortex core region [71,72]; these were interpreted as evidence of bound quasiparticles in a discrete energy state that theoretically is allowed only for a nodeless gap, noting that *d*-wave pairing symmetry would yield a continuum in quasiparticle states [23]. Coexisting hole and electron carriers found in Hall effect studies [73] have recently been identified with the two spatially separated electronic bands involved in high-$T_C$ pairing; manifestations of superconductivity is considered to have dependence on relative band sensitivities in experiment as discussed in [74].

*5.4. Magnetic field penetration at superconductor surfaces*

As mentioned in the introduction, Meissner states at the surfaces of crystals have been probed by various methods (e.g. using microwave cavities [15,16] or low energy $\mu^+$ beams [18-20]) for determining components of the magnetic penetration depth. Since magnetic fields penetrate a distance of order 0.1 μm, the results have generally been assumed as essentially equivalent to bulk measurements. Much of this work has been done on single-crystal $YBa_2Cu_3O_{7-\delta}$ and thus may be compared to the mixed-state $\mu^+$SR results summarized in section 5.3.

The low-energy $\mu^+$SR method is particularly suited for obtaining detailed depth profiles of static magnetic flux penetration in weak applied magnetic fields [18-20]. Shown in figure 8 are recent results for $\lambda_b^{-2}(T)$ obtained by Kiefl et al. for high-purity de-twinned $YBa_2Cu_3O_{6.92}$ crystals ($T_C$ = 94.1 K and $\Delta T_C < 0.1$ K by magnetometry) using Meissner effect theory and assuming a dead surface layer (10.3(5)-nm thick) to model the depth dependence of the magnetic field [18]. The filled square and circle points reproduce data presented numerically ($\lambda_b$ = 108.4 nm at T = 8 K) and graphically ($\lambda_b$ *vs.* T, figure 3), respectively, in [18]. There is a discrepancy between the two 8-K data points that falls outside the quoted experimental uncertainty (±1 nm), indicating either an erroneous datum (filled circle) or an unspecified systematic experimental error. Although penetration of fluxons in the bulk along the *a*-axis of the crystals is possible (H = 9.46 mT), inter-vortex separations are at least 0.36 μm and thus sufficiently exceed $\lambda_b$ (0.104 to 0.257 μm) for H ~ 0 theory to be applicable. The solid curve in figure 8 shows the clean *d*-wave theory (H ~ 0) from [38] drawn through the limiting points given in [18] (denoted by the arrows). The dotted curve is a fit of the modified two-fluid formula, $\lambda_b^{-2}(T) = \lambda_b^{-2}(0) [1 - (T/T_C)^\alpha]$ with $\lambda_b(0) = 108.2(6)$ nm, $\alpha = 2.1(1)$, and $T_C = 88.7(1.2)$ K (adopting the square-symbol datum as the correct one for T = 8 K). The arrow at T = 0 marks the different $\lambda_b(0) = 105.5(1.3)$ nm that is obtained in [18] by linear extrapolation (noted to be sensitive to choice of fitting points).[13] The nearly quadratic ($\alpha \approx 2$) form represented by the modified two-fluid model fit may be indicative of *d*-wave pairing in the presence of strong scattering from impurities [75], although these would be of unexplained origin in such an interpretation, owing to the high-purity of the crystals. Even though it appears that the modified two-fluid formula yields the better fit, the F-distribution test ($\chi^2_{d-wave} / \chi^2_{2-fluid} \approx 1.4$) indicates that these two model functions are statistically nearly indistinguishable.

---

[12] Such findings are proof in principle for the existence of nodeless symmetry; other studies showing more V-shaped spectra and finite zero-bias tunnelling are noted in [24].
[13] The linear extrapolation method used in [18] made no assumption regarding pairing state.

Noting that the low-energy $\mu^+$SR data yield an effective $T_C$ that is at least 5 K lower than the bulk and that a dead layer is involved in the data analysis, it is apparent that superconductivity near the surfaces of these $YBa_2Cu_3O_{6.92}$ crystals differs significantly from that in the bulk. This interpretation is consistent with $\alpha < 4$ and an unexpectedly low anisotropy, $\lambda_a/\lambda_b = 1.19$ as noted in [18]. In a previous low-energy $\mu^+$SR study of a *c*-axis oriented epitaxial $YBa_2Cu_3O_{7-\delta}$ film it was proposed that the similarly observed "dead layer" (8-nm thick) is produced by surface roughness (with a mean thickness of 5 nm rms determined from atomic force microscopy) together with a non-superconducting surface (of a few nm) [19]. In the roughness model that was proposed, Meissner screening currents become sensitive to the larger c-axis component of $\lambda_c$ (from section 5.2, $\lambda_c/\lambda_{ab} \approx 5.8$), owing to undulations of current flow in and out of the *a-b* plane.

Microwave techniques have been used to probe the dynamic response of superconducting diamagnetic screening induced at surfaces, obtaining the penetration depth in the form $\Delta\lambda = \lambda - \lambda_0$, where $\lambda_0$ is an undetermined baseline. Appreciation for these techniques stems from the relatively high precision obtainable for the temperature dependence of $\Delta\lambda$ (with signal-to-noise improved relative to $\mu^+$SR), albeit absent a determination of an absolute $\lambda$. Studies of *c*-axis oriented $YBa_2Cu_3O_{7-\delta}$ films ($T_C$ = 90 K, $\lambda_{ab}(0)$ = 140 nm) had shown an exponential form in the temperature dependence at low T [76]. On the other hand surface resistance measurements on high quality crystals (i.e. homogeneous, $\delta = 0.05$, $T_C$ = 93 K, $\Delta T_C = 0.2$ K from specific heat jump) were found to be consistent with $\Delta\lambda(T) \propto T^2$ [77]. The reanalysis presented in [77] of data similar to [76] had deduced that the $T^2$ form also applies to the epitaxial films. However, statistical analysis presented in [25] showed that $T^2$ dependence, corresponding to $\alpha \approx 2$, ought not apply to a crystal in the bulk. Later microwave surface impedance measurements on high quality untwined crystals have found the more ubiquitous result $\Delta\lambda(T) \propto T$ [15], corresponding to $\alpha \approx 1$. Differences exhibited among high quality (in the bulk) crystals, as indicated by quadratic ($\alpha \approx 2$) or linear ($\alpha \approx 1$) forms in dynamic measurements of the penetration depth, suggest that superconductivity at the surfaces is strongly affected by materials differences. In the context of the *d*-wave gap symmetry interpretation, such surfaces vary from being clean ($\alpha \approx 1$) to strongly scattering ($\alpha \approx 2$).

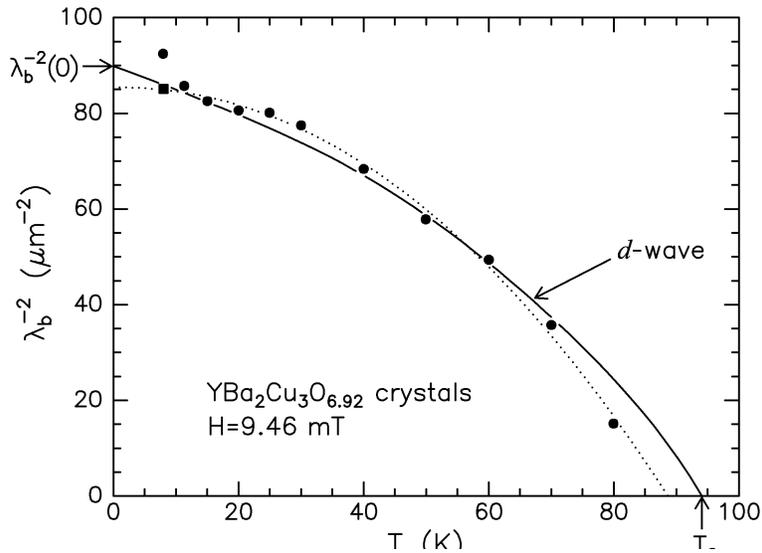

**Figure 8**. Meissner-effect b-axis penetration depth data for high-quality $YBa_2Cu_3O_{6.92}$ crystals given graphically (filled circles) and numerically (filled square) in [18]. Arrows point to T=0 and T=$T_C$ limits as given in [18]. Dotted curve is fit to the modified two-fluid model (see text). Solid curve is *d*-wave theory [38].

A number of works have proposed alternative interpretations of linear and quadratic temperature dependences in the penetration depth. Roddick and Stroud considered classical and quantum models to demonstrate theoretical possibility that thermal fluctuations of the phase of the order parameter produce a linear-in-T correction in the penetration depth of the magnitude observed [78]. It was concluded fluctuations obscure the intrinsic gap symmetry and that observations of linear-in-T forms are insufficient to demonstrate nodes in the gap. A changeover from linear to higher order temperature dependence must occur in $\Delta\lambda(T)$ in the limit $T\to 0$, based purely on laws of thermodynamics, discussed in [79,80,81], which is not found in $\alpha \approx 1$ microwave data.

Quadratic temperature dependence in the penetration depth has been observed in $YBa_2Cu_3O_{7-\delta}$ epitaxial films and $\kappa$–[BEDT-TTF]$_2$Cu[NCS]$_2$ crystals, which was associated with crystal defects that act as sites for nucleating vortex-antivortex pairs [82]; in this interpretation the exponent $\alpha \approx 2$ derives from vortex pinning forces varying with the thermodynamic critical field, $H_C(T) \sim 1 - (T/T_C)^2$. Various indicators of extrinsic surface effects have been discussed in [76], where it was noted that residual surface resistance and far infrared absorption indicate an excess normal component (in two-fluid model) or weak link behaviour; these observations were considered to be evidence for dependence of surface impedance on extrinsic materials defects. The penetration depth is derived from surface impedance measurements by extracting the superconducting kinetic inductance that is commingled with phase modulations associated with the anomalous electrodynamic dissipation component [23]. This component was of interest in recent studies of overdoped $YBa_2Cu_3O_{6.99}$ crystals, where the microwave impedance was found to contain non-Drude conductivity [83].

Evidently, the penetration depths of $YBa_2Cu_3O_{7-\delta}$ derived from Meissner-effect studies display much weaker temperature dependences ($\alpha \approx 1-2$) than in most mixed-state studies ($\alpha \approx 4$), exceptions noted in section 1. The plausible rationale is that manifestations of high-$T_C$ superconductivity at surfaces differ from those in the bulk, such as was considered in [40]. In the Meissner effect, supercurrents flow over macroscopic-scale distances ($\sim 10^{-1}$ cm crystal width), whereas in the mixed state the distances are microscopic (inter-vortex spacings $\sim 10^{-5}$ cm at 0.2 T). Consequently, Meissner-effect studies have inherent sensitivity to surface flatness and continuity of superconducting state in the surface sheath. For example, the extraordinary high quality of the crystals used in the low-energy $\mu^+$SR Meissner-effect experiments of [18] seems not to extend to the crystal surfaces. Moreover, dynamic probes (microwaves) have differing sensitivity to surface texture and roughness effects than static probes (low-energy $\mu^+$SR), even though the best results from both types of experiments employ similarly high purity crystals. In the face of these uncertainties, the substantial body of evidence from tunnelling and $\mu^+$SR pointing to a nodeless gap in $YBa_2Cu_3O_{7-\delta}$ (and other high-$T_C$ compounds) must be taken into consideration in any serious determination of the true high-$T_C$ pairing state.

*5.5. Inconsistencies within the complex gap model*

Alternative gap symmetry theory has been proposed to explain the temperature inflection features observed in the axial components $\sigma_a$ and $\sigma_b$ for $YBa_2Cu_3O_{7-\delta}$ [37] and $YBa_2Cu_4O_8$ [40] crystals and the planar component $\sigma_{ab}$ for the $La_{1.83}Sr_{0.17}CuO_4$ crystal [41], which were modelled after a two-gap interpretation of the phonon superconductor $MgB_2$ [84,85]. The temperature dependence of these $\sigma$ parameter components appear to be fitted quite reasonably using two superconducting gap functions with *d* and *s* symmetries [37,40,41]. In polycrystalline $MgB_2$ a strong and non-monotonic field dependence in $\sigma(T\to 0, H)$ is observed by $\mu^+$SR, which was attributed to pinning effects (by noting similar observations in cuprates, e.g. [4]) [85] while also considering two superconducting gaps (conflict between these interpretations, i.e. extrinsic pinning *vs.* intrinsic gap properties, being unaddressed) [85,86]. As noted in section 4, these signatures of pinning exist also in the data for the cuprate crystals and as such they tend to refute the suppositions of multiple-gap models. Moreover, the secondary component with *s*-wave symmetry obtained by fitting a two-gap model contains fundamental inconsistencies in physical properties. The gap deduced for this component is larger for $La_{1.83}Sr_{0.17}CuO_4$ than it is for $YBa_2Cu_3O_{7-\delta}$ (1.57 *vs*. 0.71 meV), even though $T_C$ is substantially lower (36.2 *vs*. 91.2 K). The secondary component also appears strongly quenched at H ~ 0.1 T for $La_{1.83}Sr_{0.17}CuO_4$, whereas it persists to the highest field for $YBa_2Cu_3O_{7-\delta}$. A larger gap is generally associated with a higher $T_C$, and accordingly a higher critical field. Since the two-gap model applied to $La_{1.83}Sr_{0.17}CuO_4$ and $YBa_2Cu_3O_{7-\delta}$ produces trends in opposition to logical expectations, it's fair to doubt its applicability to the cuprates. An opposite trend is also produced for $YBa_2Cu_4O_8$, where the secondary gaps in the *ab*-plane are on average 43% larger than for $YBa_2Cu_3O_{7-\delta}$, even though $T_C$ is significantly lower (79.9 *vs*. 91.2 K) [40]. While $T_{C0}$ = 104 K for $YBa_2Cu_4O_8$ (under hydrostatic pressure), the measured $T_C$ values (whether optimal or non-optimal) generally scale with the gap magnitude and vice-versa.

The data for the $YBa_2Cu_4O_8$ crystal were obtained at low field (H = 0.015 T), where pinning induced disorder of the fluxon lattice tends to be most pronounced (see e.g. [29]). Since the two-gap model finds an *s*-wave component gap for $YBa_2Cu_4O_8$ that is larger than that for $YBa_2Cu_3O_{7-\delta}$, where the putative *s*-wave component persists to 0.64 T, it is logical to anticipate that it would surely persist to the intermediate fields used in the powder experiments, H = 0.2 and 0.35 T (powders A [52] and B [49], respectively, in figure 4). Consequently, it is reasonable to conclude that the inflection feature is a peculiarity of the crystals, given its absence in the powder data.

As applied in all three cases, the multiple gap scenario proposed for $La_{1.83}Sr_{0.17}CuO_4$ [41] ensues by minimizing the importance of pinning in the data and adopting from prior work on $MgB_2$ [85] a model function in which σ is taken to be a linear combination of *s*-wave and *d*-wave terms. Although a linear combination model may correctly calculate $\lambda^{-2}$, it has been argued that a non-linear formulation is more appropriate for calculating the second moment [86]. Nevertheless, this two-gap fit to the data attributes the strong field dependence in $\sigma_{ab}(T\rightarrow 0,H)$ to an *s*-wave component [41], for which an extremely small upper critical field is required, estimated as $H_{c2}^{(s)}(0)$ ~ 0.2 T from equation (9). Since this *s*-wave component comprises a substantial fraction of $\sigma(T\rightarrow 0,H\rightarrow 0)$, the result implies a fluxon core area that is an order of magnitude larger than is generally understood for cuprate superconductors [14]. A non-linear refinement of the model is not likely to materially remedy this inconsistency.

## 6. Conclusion

We have considered the problem regarding the gap function symmetry derived from mixed-state $\mu^+$SR data for three high-$T_C$ compounds from the perspective of three main schools of thought: The first asserts that σ(T) shows evidence of a single *d*-wave gap (linearity-in-T model) [14,36]; the second interprets the inflection in σ(T) as evidence of multiple gaps (*s* + *d* in the basal plane and *s*-wave along the c-axis) [37,40,41]; and the third proposes that σ(T,H) data can be explained by temperature-activated fluxon depinning and disorder [25,26]. Only the third interpretation is consistent with the four key observations: (1) A qualitatively different behaviour of σ(T,H) in single-crystal samples, when compared to non-oriented sintered powder samples of the same superconductor, as well as an irreproducible linearity-in-T response in crystals; (2) A differing of results for the same crystal and apparatus; (3) A non-monotonic variation of σ(T→0,H) with H; and (4) A transition temperature which appears to decrease with applied field for H << $H_{C2}$(T). Thus the third interpretation, arrived at in this paper, is the only viable one.

A careful analysis of the temperature and field dependences of σ(T,H) for crystal and randomly-oriented powder samples of $YBa_2Cu_3O_{7-\delta}$, $YBa_2Cu_4O_8$ and $La_{2-x}Sr_xCuO_4$ (x ~ 0.15 − 0.17) yields clear and unambiguous evidence of the precise textbook behaviours expected for fluxon lattices in highly anisotropic superconductors [2,33]. Given the strong intrinsic pinning for fluxons aligned parallel to the a-b plane [64], thus restricting vortex movement along the *c*-axis, the *c*–axis component $\sigma_c$(T) would best reflect the true symmetry of the underlying gap function, which is found to be nodeless and consistent with *s*-wave, in both $YBa_2Cu_3O_{7-\delta}$ and $YBa_2Cu_4O_8$. Temperature-activated fluxon depinning, present in axial components $\sigma_a$ and $\sigma_b$ as well as in the three planar components, obscures the underlying gap function.

We have also touched on how sample inhomogeneities arising from non-optimization might introduce a factor of $[1+\xi/\ell(T)]^{-1}$ multiplying $\sigma(T,H)$ that can mimic *d*-wave pairing. Empirically, this corresponds to finding $\alpha < 4$ (insets to figures 2 and 4; figure 8). Since the pairing state cannot change with the degree of non-optimization, such phenomena must also be considered in any interpretation of $\sigma(T,H)$ in (non-ordered) oxygen-deficient $YBa_2Cu_3O_{7-\delta}$ (see inset of figure 2), ambient pressure $YBa_2Cu_4O_8$ data (see inset of figure 4), non-stoichiometric $La_{2-x}Sr_xCuO_4$, and in other non-optimal high-$T_C$ compounds.

The fact that the anomalous inflection in $\sigma(T)$ is absent in all of the non-oriented sintered powder data (as well as in all of the crystal data displayed in [14,36]) proves it to be extrinsic to the superconductivity, and not reflective of a complex gap function associated with the superconducting hole condensate. This simple physical explanation can be applied to all related $\mu^+SR$ data, where the various fluxon effects and dependences are determined by the pinning potential distribution [2,65] and $\Gamma$. Indeed, far from providing evidence of multiple gaps, the data on crystals instead provide corroborative evidence of a single dominant gap symmetry for the superconducting hole condensate. From F-distribution statistical analysis the results for $YBa_2Cu_3O_{7-\delta}$ (crystals and powders) are consistent with a nodeless strong-coupled ground state symmetry; a similar finding is obtained the case of $YBa_2Cu_4O_8$, where a *d*-wave gap is found to have low statistical significance, except for the powder-A specimen, where likelihood of a noded gap appears greater. Similar analysis of $La_{2-x}Sr_xCuO_4$ confirms a nodeless gap for the powder data; the crystal data show a better fit with a nodeless gap, although a *d*-wave gap is statistically possible. Hence, from the perspective of $\mu^+SR$ studies of the mixed state, the gap associated with the superconducting hole condensate in the bulk of these high-$T_C$ superconductors is nodeless in those cases where analysis produces statistical significance (consistent with strong-coupled *s*-wave pairing) for supercurrent flows in both the basal plane and also along the hard-axis (*c*-axis). Finally, we considered evidence from magnetic field penetration profiling of the Meissner state using $\mu^+SR$, which suggests superconductivity in surface sheaths of crystals differs from the bulk, and note that many of the experimental indicators of *d*-wave pairing symmetry involve surfaces or interfaces. Our analysis of the $\mu^+SR$ data considered herein in fact shows no statistically significant evidence for *d*-wave pairing of the superconducting hole condensate.

## Acknowledgements

We are grateful to Physikon Research Corporation (Project No. PL-206) and the New Jersey Institute of Technology. We also appreciate helpful discussions with Prof. John D. Dow and Prof. Alan J. Drew. Publication on this work has appeared [87].